\documentclass[12pt]{article}
\usepackage{epsfig}
\usepackage{graphicx}

\newcommand{\be}{\begin{equation}}
\newcommand{\dd}{\displaystyle}
\newcommand{\ee}{\end{equation}}
\newcommand{\bea}{\begin{eqnarray}}
\newcommand{\eea}{\end{eqnarray}}

\newcommand{\nn}{\nonumber}
\newcommand{\de}{\partial}
\begin{document}
\hfill{\bf BARI-TH 442/02}\par \hfill{\bf FIRENZE
DFF-390/06/02}\par
 \hfill{\bf UGVA-DPT-2002 07/1101}

\hfill$\vcenter{
  \hbox{\bf}
 \hbox{\bf } }$
\begin{center}
{\Large\bf\boldmath {Phonons and gluons in the crystalline color
superconducting phase of QCD}} \rm \vskip1pc {\large R.
Casalbuoni$^{a,b}$, E.Fabiano$^{c,d}$, R. Gatto$^e$,
 M. Mannarelli$^{c,d}$ and
 G. Nardulli$^{c,d}$}\\ \vspace{5mm} {\it{
$^a$Dipartimento di
 Fisica, Universit\`a di Firenze, I-50125 Firenze, Italia
 \\
 $^b$I.N.F.N., Sezione di Firenze, I-50125 Firenze, Italia\\ $\dd
^c$Dipartimento di Fisica,
 Universit\`a di Bari, I-70124 Bari, Italia  \\$^d$I.N.F.N.,
 Sezione di Bari, I-70124 Bari, Italia\\
 $^e$D\'epart. de Physique Th\'eorique, Universit\'e de Gen\`eve,
 CH-1211 Gen\`eve 4, Suisse }}
 \end{center}
 \begin{abstract}
The High Density Effective Theory formalism is used to calculate
the low energy properties of the  phonons and gluons in the
Larkin-Ovchinnikov-Fulde-Ferrell (LOFF) phase of two-flavor QCD.
The effective phonon Lagrangian for the cubic crystal structure,
which is favored according to a recent study, depends, at the
second order in the derivatives,  on three parameters which we
calculate in this paper. We also compute for soft momenta the
effective lagrangian for the gluons of the unbroken $SU(2)_c$
group, both for a single plane wave structure and for the cube.
In both cases the Meissner gluon mass vanishes as in the case of
complete isotropy; on the other hand there is a partial Debye
screening due to the existence of blocking regions on the Fermi
spheres. For the single plane wave structure the gluon dielectric
tensor is non isotropic, while it is isotropic for the cubic
crystal, in spite of the intrinsic anisotropy of the structure.

 \end{abstract}

\section{Introduction}
The study of the color superconducting phases of QCD
\cite{barrois} \cite{alford} has recently received much attention
and is  at the moment one of the most studied frontier problem of
QCD \cite{review}. Cold QCD with three degenerate flavors, for
very large values of the quark number chemical potential $\mu$,
is in the so-called color-flavor locked (CFL) phase \cite{arw}.
In this phase,
 pairs of quarks of opposite momenta, each of them lying near
the Fermi surface, condense according to the BCS pairing scheme
known from ordinary superconductivity.

The CFL phase is expected to persist also below asymptotic values
of $\mu$,  when the flavor degeneracy is broken, with the strange
quark having a larger mass than the lighter quarks. In absence of
electrons, the CFL phase is electrically neutral. It has the same
number of $u$, $d$, and $s$ quarks, which corresponds to the
lowest energy for the Cooper pairs.

When the value of $\mu$ is further decreased, one expects that
pairing between quarks of non vanishing total momentum becomes
possible. Each quark momentum will lie near its Fermi surface
(Fermi spheres will now be of different radii)
\cite{abr,bkrs,casa,leib,reviewsloff}. Closeness to their Fermi
surfaces for both quarks in the pair will correspond to a lower
free energy. Such a possibility for the quark pairing leads to the
suggestion of a  crystalline color superconductivity
\cite{abr,bkrs,casa,leib,reviewsloff}. We assume the existence of
this phase and we will limit our analysis, as in the existing
literature, to two flavors, with $\delta\mu\not=0$ ($\delta\mu$
the difference in the quark chemical potentials).

In certain condensed matter systems (ferromagnetic systems with
paramagnetic impurities) \cite{loff} the  possibility of fermion
pairing of non vanishing total momentum had already been
discussed. The name LOFF phase is therefore commonly used, formed
from the initials of the authors of ref. \cite{loff}.

Translational and rotational invariance are spontaneously broken
within a LOFF phase. The gaps are expected to vary according to
some crystalline structure. The determination of such a structure
is obviously of the uppermost importance and this problem is
discussed and approximately solved by  Bowers and Rajagopal
\cite{bowers} in a recent paper. To understand the crystalline
superconductivity one has indeed to know the space dependence of
the Cooper pair condensate. A simple plane wave ansatz had been
used for a preliminary understanding of the possible dynamics. The
plane wave has the finite momentum of the Cooper pairs, on the
assumption that all the possible Cooper pair momenta choose a
unique direction. These studies (see  \cite{abr} and \cite{loff})
already lead to the conclusion  that LOFF pairing is favored,
over BCS pairing or absence of pairing, within a certain window
in $\delta\mu$. The window starts with a first order phase
transition to the crystalline phase. It ends with a second order
transition where translation invariance is restored. The window
for QCD with two flavors had been found in ref. \cite{abr} by
modeling quark interactions via four-linear Fermi couplings among
quarks and for the case of a plane-wave condensate structure. A
much wider window is found however in a model based on gluon
exchange \cite{leib}. This follows essentially from the fact that
gluon exchange favors forward scattering, rather than the s-wave
scattering typical of the Fermi couplings. The effective low
energy phonon lagrangian  for the case of the single plane wave
structure has been discussed in \cite{casa}. Since translation
invariance is only partially broken one phonon is present.

A comprehensive study of the different possible crystal
structures for the crystalline color superconducting phase of QCD
is presented in \cite{bowers} where the authors perform a
Ginzburg-Landau expansion in the vicinity of the phase
transition. Instead of the single plane wave ansatz they now
introduce a multiple plane-wave ansatz and find that a
face-centered cubic crystal is  favored. This structure is built
up by eight plane waves directed to the corners of a cube.   At
least qualitatively it seems true that the  cubic crystal remains
the favored structure also in complete QCD (although not perhaps
at very asymptotic densities). It is important to note that for
the cubic crystal structure three phonons should be present since
translational invariance is completely broken.

The crystalline color superconductivity has a relevant place in
the QCD phase diagram and as such it deserves to be studied and
to be known in its main features. Moreover it can play a role in
the physics of compact stellar systems where, within the core, a
crystalline phase might be reached.  The authors of ref.
\cite{abr} had already pointed out that, if the compact star is
rotating, the ensuing rotational vortices may become pinned and
as such play a role in the formation of glitches. This subject is
certainly interesting, but we shall not discuss it here (for a
review see ref. \cite{beppe}).

We have already mentioned that crystalline superconductivity had
initially been proposed in condensed matter physics. The QCD
studies, in particular also the present one, may be translated
into corresponding properties for condensed systems, at least for
the phonon properties. We mention in particular the field of
atomic systems, gases of fermionic atoms below their Fermi
temperature \cite{screck}, \cite{holland}. An advantage of testing
the ideas of crystalline superconductivity on these systems is
that their parameter space is less rigid than for QCD. For
example  see  ref. \cite{combescot}.

The plan of the present work is as follows. In section 2 we first
briefly review the formalism we will use in this paper, i.e. the
High Density Effective Theory (HDET). This effective theory is
based on the limit $\mu\to\infty$ and uses velocity dependent
fields in the same spirit of the Heavy Quark Effective Theory
(HQET) \cite{HQET} where the limit $m_Q\to \infty$ is taken. It is
discussed in \cite{hong, beane, cflgatto} and  more extensively
in ref. \cite{rivistabeppe}, where the method and the notations
are explained in more detail.  In section 3 we discuss the low
energy dynamics of the phonons. In section 4 we calculate the
parameters of the effective phonon lagrangian. In section 5 we
study the properties of the gluons of the unbroken $SU(2)_c$  for
the case of the plane wave structure and in section 6 we perform
the same calculation for the  LOFF condensate with the cubic
symmetry. The application to the cubic crystal structure requires
some group theoretical definitions which  are summarized in
Appendix 1. In Appendix 2 and 3 we report  some details of the
calculations.

 \section{Heavy Density Effective Theory
in the LOFF phase\label{hdet}}

We begin with the effective lagrangian of HDET for gluons and
ungapped massless quarks. If ${\cal L}_{0}$ is the free quark
lagrangian and ${\cal L}_{1}$ represents the coupling of quarks
to one gluon, the high density effective lagrangian can be
written as \be {\cal L}_D={\cal L}_{0}+{\cal L}_1+{\cal L}_2\ +\
(L\to R)\,.\label{dirac} \ee Let us begin with the first two
terms, i.e. \bea {\cal L}_{0}& =& \sum_{\vec v}
\Big[\psi_+^\dagger iV\cdot
\partial\psi_+\ +\
\psi_-^\dagger i\tilde V \cdot \partial\psi_-\Big]\ ,\\
{\cal L}_1& =&i\,g\, \sum_{\vec v} \Big[\psi_+^\dagger iV\cdot
A\psi_+\ +\ \psi_-^\dagger i\tilde V \cdot A\psi_-\Big]\
\label{l1}\ .\eea The velocity dependent left-handed field
$\psi_+\equiv \psi_{\vec v}$ is the positive energy part in the
decomposition \be\psi(x)=\int\frac{d\vec v}{4\pi} e^{-i\mu v\cdot
x}
 \left[\psi_{\vec v}(x)+\Psi_{\vec v}(x)\right]\ \ee
while $\Psi_{\vec v}$ is the negative energy part which has been
integrated out. Here \be \psi_{\vec v}(x)=e^{i\mu v\cdot
x}P_+\psi(x)=\int_{|\ell|<\delta}\frac{d^4\ell}{(2\pi)^4}e^{-i\ell\cdot
x}P_+\psi(\ell)\ee and \be \Psi_{\vec v}(x)=e^{i\mu v\cdot
x}P_-\psi(x)=\int_{|\ell|<\delta}\frac{d^4\ell}{(2\pi)^4}e^{-i\ell\cdot
x}P_-\psi(\ell)\,.\ee $P_\pm$ are projectors defined by \be
 P_\pm= \frac{1}2 \left(1\pm{\vec\alpha\cdot\vec v}\right)
  \ .\ee The cut-off $\delta$ satisfies $\delta\ll\mu$ while being
  much larger that the energy gap.  These equations correspond to the following
decomposition of the quark momentum: \be p^\mu=\mu
v^\mu+\ell^\mu\, \label{4.1.dec} \ee where \be v^\mu=(0,\vec v )
\ee and $|\vec v|=1$. We also use $ V^\mu=(1,\,\vec v)$ ,$\tilde
V^\mu=(1,\,-\vec v)$.

The other velocity-dependent field $\psi_-\equiv \psi_{-\vec v}$
is obtained from $\psi_+$ by the change $\vec v \to - \vec v$.
Therefore the average over the Fermi velocities is defined as
follows: \be \sum_{\vec v}=\int\frac{d\vec v}{8\pi}\ .
\label{sumvel} \ee The extra factor $\dd \frac 1 2$
  is introduced because, after the introduction of the
field with opposite velocity $\psi_-$, one doubles the degrees of
freedom, which implies that the integration is only over half
solid angle.

The effective lagrangian (\ref{dirac}) is obtained in the limit
$\mu\to\infty$ \cite{hong,beane,cflgatto,rivistabeppe}. Therefore
${\cal L}_0$ and ${\cal L}_1$ do not contain the chemical
potential because the dependence on $\mu$ has been extracted from
the quark field operators.  We also note that one has a single
sum over the velocities, because the quark fields
 are  evaluated at the same Fermi velocity; as a matter of fact
 the  off-diagonal terms
are  cancelled by the rapid oscillations of the exponential factor
in the $\mu\to\infty$ limit. This point will be discussed in a
more detailed way in the sequel.

Besides ${\cal L}_0$ and  ${\cal L}_1$ eq.(\ref{dirac}) contains
the term \be {\cal L}_2=-\sum_{\vec v}\,P^{\mu\nu}\, \Big[
\psi_+^\dagger\frac{1}{2\mu+i\tilde V\cdot D}D_\mu D_\nu \psi_+ +
\psi_-^\dagger\frac{1}{2\mu+iV\cdot D}D_\mu D_\nu
  \psi_-\Big]\ .\label{l2}
  \ee ${\cal L}_{2}$ is a non local lagrangian
 arising when one integrates over the  $\Psi_{\vec v}$
 degrees of freedom in the functional integration.
  It contains couplings of two quarks to any number of gluons
  and
gives contribution to the gluon
 Meissner mass (see below). We have put \be P^{\mu\nu}=g^{\mu\nu}-\frac 1
2\left[V^\mu\tilde V^\nu+V^\nu\tilde V^\mu\right]\ .\label{219}\ee

 This construction  is  valid for any theory
describing massless fermions at high density provided one excludes
 degrees of freedom  far from the Fermi surface. Let us
 now  examine
  the  modification in the formalism in the  presence of
a LOFF condensate; to begin with we consider the approximation of
one plane wave. The space dependence of the crystalline condensate
in this case is \be \Delta(\vec x)=\Delta\,\exp\{2i\vec q\cdot\vec
x\}\ . \label{1}\ee
 The effect of the non vanishing vacuum
expectation value can be taken into account by adding to the
lagrangian the term: \be {\cal
L}_\Delta=-\frac{\Delta}2\,\exp\{2i\vec q\cdot\vec
x\}\,\epsilon_{\alpha\beta 3} \epsilon_{ij}\psi_{i\alpha}^T(x)C
\psi_{j\beta}(x)\,-(L\to R)+{\rm h.c.}\ .\label{ldeltaeff}\ee

 In order to introduce velocity dependent positive energy
fields $\psi_{\vec v_i;\,i\alpha}$ with flavor $i$ we decompose
both fermion momenta as in (\ref{4.1.dec}) and we have: \bea {\cal
L }_\Delta&=&-\frac{\Delta} 2\, \,\sum_{\vec v_i,\vec v_j}
\exp\{i\vec x\cdot\vec\alpha(\vec v_i,\,\vec v_j,\,\vec q
)\}\epsilon_{ij}\epsilon_{\alpha\beta 3}\psi^T_{-\,\vec
v_i;\,i\alpha}(x)C \psi_{-\,\vec v_j;\,j\beta}(x)\cr && -(L\to
R)+{\rm h.c.}\ ,\label{loff6}\eea where \be\vec\alpha(\vec
v_i,\,\vec v_j,\,\vec q)=2\vec q-\mu_i\vec v_i-\mu_j\vec
v_j\,.\ee Let us define
\be\mu=\frac{\mu_1+\mu_2}{2}~,~~~~~~~~~\delta\mu=\,-\,\frac{\mu_1-\mu_2}{2}
.\label{dec2}\ee We will consider large values of $\mu$ (with
$\delta\mu/\mu\to 0,~ q/\delta\mu\to 0$ and $\delta\mu/q$ fixed),
but below the value where the CFL phase sets in. To give  our
procedure a mathematical meaning, we consider a smeared amplitude
as follows:
 \be\lim_{\mu\to\infty}\exp\{i\vec r\cdot\vec\alpha(\vec
v_1,\,\vec v_2,\,\vec q)\} \equiv \lim_{\mu\to\infty}\int d\vec
r^{\,\prime}\, \exp\{i\vec r^{\,\prime}\cdot\vec\alpha(\vec
v_1,\,\vec v_2,\,\vec q)\}g(\vec r,\vec r^{\,\prime}) \
.\label{1bis}\ee In \cite{effLOFF2} we  assumed the following
 smearing function: \be g(\vec r,\vec
r^{\,\prime})=g_I(\vec r-\vec
r^{\,\prime})=\prod_{k=1}^3\theta_{\Delta\ell}(r_k-r^\prime_k)
\label{gI}\ee where \bea
 \theta_{\Delta\ell}(r-r^\prime)&=&\frac
1{\Delta\ell}\left[\theta\left(r^\prime-r+\frac{\Delta\ell}{2}\right)-
\theta\left(r^\prime-r-\frac{\Delta\ell}{2}\right)\right]=\cr&&\cr&&\cr&=&{\dd\Bigg
\{ } \begin{array}{cc} {\frac{1}{\Delta\ell}} & \textrm{~~for~~}
|r^\prime-r| < \frac{\Delta\ell}{2} \cr&\cr
0 & \textrm{elsewhere} \\
\end{array}
\eea  ($\theta$ the Heaviside function). In the present paper,
for reasons to be discussed below, we will choose \be g(\vec
r,\vec r^{\,\prime})=g_{II}(\vec r-\vec
r^{\,\prime})=\prod_{k=1}^3 \frac{\sin\left[\dd\frac{\pi
q(r_k-r^\prime_k)}{R}\right]}{\pi(r_k-r^\prime_k)}\,.
\label{gII}\ee

We evaluate (\ref{1bis}) in the $\mu\to\infty$ limit by taking
$\vec q$ along the $z-$axis, and using the following identity: \be
\int d^3{\vec r}^{\,\prime} \exp\{i\vec
r^{\,\prime}\cdot\vec\alpha\}g_{II}(\vec r-{\vec r}^{\,\prime})=
\exp\{i\vec
r\cdot\vec\alpha\}\left(\frac{\pi}R\right)^3\delta_R^3\left(\frac{\vec\alpha}{2q}\right)\ee
where \be \delta_R(x)= {\dd\Bigg \{ }
\begin{array}{cc} {\dd\frac{R}{\pi}} & \textrm{~~for~~}
|x| < \dd{\frac{\pi}{2R}} \cr&\cr
0 & \textrm{elsewhere} \\
\end{array}
\label{deltaII}\ee For the components $x$ and $y$ of $\vec\alpha$
we get \be |(\mu_1 v_1+\mu_2 v_2)_{x,y}|<\frac{\pi q}R\,,\ee i.e.
approximately (for $\delta\mu\ll\mu$) \be |( v_1+
v_2)_{x,y}|<\frac{\pi q}{R\mu}\,.\label{25b}\ee From this in the
high density limit it follows \be\vec v_1= -\vec v_2+{\cal
O}(1/\mu)\,.\ee More precisely, if $\theta_1$ and $\theta_2$ are
the angles of $\vec v_1$ and $\vec v_2$ with respect to the
$z$-axis one gets \be
\theta_1=\theta_2+\pi+2\frac{\delta\mu}{\mu}\tan\theta_2\,.\ee
For the $z$ component we get \be \alpha_z=2qh(\cos\theta_2)\,.\ee
The expression for $h(x)$ is  \cite{effLOFF2}: \be h(x)=1+\frac
{x\mu_2} {2q}\left(-1+\sqrt{1-\frac{4\mu\,\delta\mu}{\mu_2^2\,
x^2}}\right)\,. \label{29}\ee Neglecting corrections of order
$\delta\mu/\mu$ one finds \be
h(x)=1-\frac{\cos\theta_q}x\,,~~~\cos\theta_q=\frac{{\delta\mu}}q\,.\label{hdix}\ee

The two factors $\pi/R$ arising from the $x$ and $y$ components
are absorbed into a wave function renormalization of the quark
fields, both in the kinetic and in the gap terms. As for the $z$
component one remains with the factor \be
\frac{\pi}{R}e^{i2qhz}\,\delta_R[h(\vec v\cdot\vec
n)]\approx\frac{\pi}{R} \delta_R[h(\vec v\cdot\vec
n)]\label{dr}\ee where \be\vec n=\vec q/q\ \ee in the gap term,
whereas for the kinetic term we get a factor of 1.

As we noted already, in ref. \cite{effLOFF2} we used as a smearing
function $g_I$ (see eq. (\ref{gI})) and  $\delta_R$ turned out to
be given by\be \delta_{\hat R}(x)= {\dd\frac{\sin({\hat R}x)}{\pi
x}}\label{deltaI},~~~ \hat R=q|\Delta \ell|\,.\ee

 Both eqns. (\ref{deltaI}) and (\ref{deltaII}) define regions
where $\delta_R\neq 0$, i.e. define domains where pairing between
the two quarks can occur; they correspond to the {\it pairing
region} (in contrast with the {\it blocking region}, where
$\delta_R=0$), using the terminology of \cite{arw}. However, given
the results of \cite{bowers}, which show that the different
sections of the pairing domain for the LOFF cubic structure are
non-overlapping, we prefer to choose the prescription (\ref{gII})
and, consequently, eq. (\ref{deltaII}) for $\delta_R$. We
explicitly observe that, owing to (\ref{hdix}), with $x=\vec
v\cdot\vec n=\cos\theta$, eq. (\ref{deltaII}) defines a 'ring' on
the Fermi sphere. Our approximation consists in putting
  $\exp[i2qhz]=1$ in eq. (\ref{dr}), owing to the presence of
  $\delta_R$ that enhances the
  domain of integration where $h=0$, for $R$ sufficiently large.

Let us discuss the range of values of $R$\footnote{Notice that
the quantity $R$ used in the present work is not the same used in
\cite{effLOFF2}.} for the plane wave ansatz in more detail. We
notice that $R=\infty$ implies the vanishing of the pairing
region. The precise value of $R$ could be fixed by the gap
equation. For instance, by using the fact that near the second
order phase transition the pairing region has an angular width of
order $\Delta/\delta\mu$ (see for a discussion \cite{bowers}),
from the definition of $\delta_R$ we find (for $\Delta\to 0$) \be
\frac{\pi}R\approx \frac{\Delta}{\delta\mu}\sqrt{\left(\frac
q{\delta\mu}\right)^2-1}\,.\ee This shows that $R\to\infty$ for
$\Delta\to 0$. In any case for the purpose of this paper we leave
$R$  as a  parameter.

In conclusion \be {\cal L }_\Delta=-\frac{\Delta} 2\, \frac\pi
R\,\sum_{\vec v}\delta_R[h(\vec v\cdot\vec
n)]\,\epsilon_{ij}\epsilon_{\alpha\beta 3}\psi^T_{\vec
v;\,i\alpha}(x)C \psi_{-\,\vec v;\,j\beta}(x)-(L\to R)+{\rm
h.c.}\label{loff6bis}\ee

 We now change the basis
for the fermion fields
 by writing \bea \psi_{+,\alpha i}&=&
\sum_{A=0}^3\frac{(\sigma_A)_{\alpha i}}{\sqrt 2}\psi_{+}^A
~~~~~~~~(i,\,\alpha=1,\,2)~\cr \psi_{+,3 1}&=&\psi_{+}^4\cr
\psi_{+,3 2}&=&\psi_{+}^5\ ,\eea where $\sigma_A$ are the Pauli
matrices for $A=1,2,3$ and $\sigma_0=1$; similar expressions hold
for $\psi^A_-$.

A different, but also useful notation
 for the fields $\psi_{+,\,\alpha i}$ uses
 a combination of $\lambda$ matrices, as follows
\be \psi_{+,\alpha i}= \sum_{A=0}^5\frac{(\tilde\lambda_A)_{\alpha
i}}{\sqrt 2}\psi_{+}^A ~.\ee The $\tilde\lambda_A$ matrices are
defined in terms of the usual $\lambda$ matrices as follows: $\dd
\tilde\lambda_0=\frac 1{\sqrt 3}\lambda_8\,+\, {\sqrt\frac 2
3}\lambda_0$; $\dd \tilde\lambda_A=\lambda_A\,(A=1,2,3)$; $\dd
\tilde\lambda_4=\frac{\lambda_{4-i5}}{\sqrt 2}$; $\dd
\tilde\lambda_5=\frac{\lambda_{6-i7}}{\sqrt 2}$.
 In this basis
 one gets
\bea  {\cal L}_{0}+{\cal L}_1+ {\cal L}_\Delta&=&\sum_{\vec
v}\sum_{A,B=0}^5 \chi^{A\dagger}\left(\matrix{iTr[\tilde
T_A^\dagger\,V\cdot D\,\tilde T_B ]& \Delta_{AB}\cr \Delta_{AB}
&iTr[\tilde T_A^\dagger\,\tilde V\cdot D^*\,\tilde T_B
]}\right)\chi^B\cr&&\cr &+& (L\to R)\ . \label{2sccomplete0} \eea
Here \be \chi^A=\frac 1{\sqrt 2}\left(\matrix{\psi^A_+\cr
C\psi^{A*}_-}\right)\label{chiA}\ee and \be \tilde
T_A=\frac{\tilde\lambda_A}{\sqrt 2}\hskip1cm(A=0,...,5)\
.\label{380}\ee where the  matrix $\Delta_{AB}$ is as follows: $
\Delta_{AB}=0$ $(A\,{\rm or} \,B=4\,{\rm or} \,5 )$, and, for
$A,B=0,...,3$:\be \Delta_{AB}=\Delta_{eff}
\left(\begin{array}{cccc}
   1& 0 & 0 & 0 \\
  0 & -1 &0 & 0 \\
  0 & 0&-1& 0 \\
  0 & 0 & 0 & -1
\end{array}\right)_{AB}\ ,\label{eq:38}
\ee having put\be\Delta_{eff}=\frac{\Delta\pi} R\delta_R[h(\vec
v\cdot\vec n)]\ . \label{deltaeff}\ee From these equations one can
derive the  propagator for gapped quarks:
 \be D_{AB}(\ell)=\frac{1}{V\cdot
\ell\,\tilde V\cdot \ell\,-\,\Delta_{eff}^2} \left(
\begin{array}{cc}
\tilde V\cdot\ell\,\delta_{AB} &  -\Delta_{AB}
\\
 -\Delta_{AB}
 &
 V\cdot\ell\,\delta_{AB}
\end{array}
\right)\label{propagatore}\ .\ee On the other hand the propagator
for the fields $\chi^{4,5}$ does not contain gap mass terms and is
given by \be D(\ell)=\left(
\begin{array}{cc}
 (V\cdot\ell)^{-1}&0\\
 0&  (\tilde V\cdot\ell)^{-1}
\end{array}
\right)~.\ee

 These formulae  can be easily generalized
to the case of the face centered cube \cite{bowers}. This is
characterized by eight $\vec q_j$ vectors  with equal length
$q=\pi/a$ and directions corresponding to the eight vertices of a
cube.

In Fig. 1 some of the symmetry axes of this cube are shown: they
are denoted as $C_4$ (the three 4-fold axes), $C_3$ (the four
3-fold axes) and $C_2$ (the six 2-fold axes). For more details on
the symmetry group of the cube see  Appendix 1 and
\cite{hamermesh}. The space dependence of the condensate
corresponding to this lattice is as follows\be \Delta(\vec
x)=\Delta\sum_{k=1}^8\,\exp\{2i\vec q_k\cdot\vec
x\}=\Delta\sum_{\epsilon_i=\pm}\exp \left\{\dd{\frac{2\pi
i}a}(\epsilon_1 x_1+\epsilon_2 x_2+\epsilon_3 x_3)\right\}\ ,
\label{8a}\ee where the eight $\vec q_j$ vectors with equal length
\be q=\pi/a\ee
 have  directions corresponding to the eight
vertices of a cube.

\begin{center}
\includegraphics*[scale=.35]{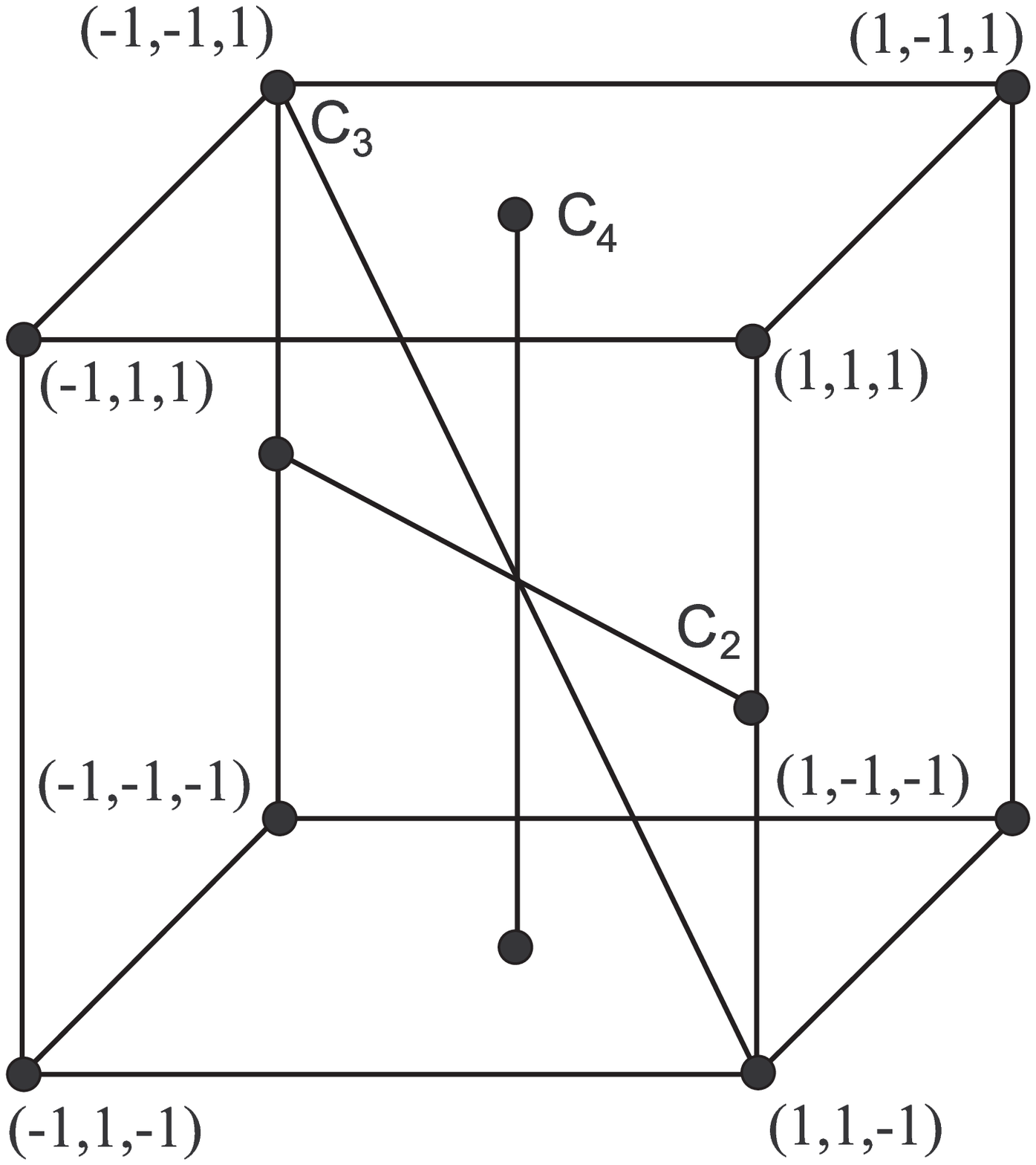}
\end{center}\noindent Fig. 1 - {\it{The figure shows  some of the
symmetry axes of the cube, denoted as  $C_2$, $C_3$ and $C_4$ (see
the appendix for details). We show also the parameterization used
for the
coordinates of the vertices.}}\\

 To describe the quark condensate we add the term: \be {\cal
L}_\Delta=-\frac{\Delta}2\sum_{k=1}^8\,\exp\{2i\vec q_k\cdot\vec
x\}\,\epsilon_{\alpha\beta 3} \epsilon_{ij}\psi_{i\alpha}^T(x)C
\psi_{j\beta}(x)\,-(L\to R)+{\rm h.c.}\ ,\label{ldeltaeff2}\ee
completely analogous to (\ref{ldeltaeff}).

Let us put \be \vec n_k=\vec q_k/|\vec q_k|\label{nk}.\ee

By the same procedure used for the plane wave condensate one has
\bea {\cal L }_\Delta&=&-\frac{\Delta} 2\sum_{k=1}^8\sum_{\vec
v}\, \frac\pi R\,\delta_R[h(\vec v\cdot\vec
n_k)]\epsilon_{ij}\epsilon_{\alpha\beta 3}\psi^T_{\vec
v;\,i\alpha}(x)C \psi_{-\,\vec v;\,j\beta}(x)\cr && -(L\to R)+{\rm
h.c.}\label{loff6b}\eea In (\ref{nk}) the eight unit vectors
defining the vertices of the cube are \bea &&\vec
n_1=\frac{1}{\sqrt 3}(+1,+1,+1),~~~\vec n_2=\frac{1}{\sqrt
3}(+1,-1,+1),~\cr &&\vec n_3=\frac{1}{\sqrt 3}(-1,-1,+1),~~~\vec
n_4=\frac{1}{\sqrt 3}(-1,+1,+1),\cr &&\vec n_5=\frac{1}{\sqrt
3}(+1+,1,-1),~~~\vec n_6=\frac{1}{\sqrt 3}(+1,-1,-1),\cr &&\vec
n_7=\frac{1}{\sqrt 3}(-1,-1,-1),~~~\vec n_8=\frac{1}{\sqrt
3}(-1,+1,-1)~ .\eea

 ${\mathcal L}_0 +
 {\mathcal L}_1 + {\mathcal L}_\Delta
 $ is still given by eqns. (\ref{2sccomplete0})-(\ref{eq:38})
but now \be\Delta_{eff}=\frac{\Delta\pi}
R\sum_{k=1}^{8}\delta_R[h(\vec v\cdot\vec n_k)]\ ;
\label{deltaeff3}\ee the quark propagator is given by
(\ref{propagatore}) with $\Delta_{eff}$ given by
(\ref{deltaeff3}).

An interesting point can be noted in connection with eq.
(\ref{deltaeff3}). This equation shows that the pairing region for
the cubic LOFF condensate is formed by eight distinct rings; each
ring is associated to one vertex of the cube and has as its
symmetry axis one of the $C_3$ axes of fig. 1. According to the
analysis of \cite{bowers} the LOFF vacuum state corresponds to a
situation where these domains have at most one common point. Given
the symmetry of the cubic structure we can limit the analysis to
one pair of rings, for example those associated to the vertices
$\vec n _1,\,\vec n_5$ in fig. 1.  The common point between these
two rings lies on the axis $C_2$ of fig. 1 and has $\dd\vec
v=\frac 1{\sqrt 2} (1,\,1,\,0).$ Since it must also belong to the
boundary of the two pairing regions we have the condition
(\ref{deltaII}):\be |h(\vec v\cdot\vec n_1)|=\frac \pi {2R }\ ,\ee
which implies \be R\,=\frac\pi{2h\left({\sqrt{2/3}}\,\right)} \
.\ee In this particular case  the approximate form for $h(x)$
(see eq. (\ref{hdix})) is not appropriate owing a numerical
cancellation. In fact for ${q}/{|\delta\mu|}\approx 1.2$,
$\mu=400~MeV$, $\delta\mu=40~MeV$, one would get $R\approx 76$,
whereas the use of the complete equation (\ref{29})
 gives \be R\approx
18\,.\label{RT}\ee

\section{Phonons in a cubic LOFF crystal}
 The condensate (\ref{8a})
breaks both translations and rotations. It is however invariant
under the discrete group $O_h$, the symmetry group of the cube.
This can be seen  by noticing that the condensate is invariant
under the following coordinate transformations \bea
&&R_1:~~~ x_1\to x_1,~~~x_2\to x_3,~~ x_3\to -x_2,\nn\\
&&R_2:~~~ x_1\to -x_3,~~~x_2\to x_2,~~~ x_3\to x_1,\nn\\
&&R_3:~~~x_1\to x_2,~~~x_2\to -x_1,~~~x_3\to x_3,\nn\\ &&I:~~~~~
x_1\to -x_1,~~~x_2\to -x_2,~~~x_3\to -x_3, \label{basic}\eea that
is rotations of $\pi/2$ around the coordinate axes, and the
inversion with respect to the origin. Since the group $O_h$ is
generated by the previous 4 elements (see Appendix 1) the
invariance follows at once.

The crystal defined by the condensate (\ref{8a}) can fluctuate and
its local deformations define three phonon fields $\phi^{(i)}$
that are the Nambu-Goldstone bosons associated to the breaking of
the translational symmetry. They can be formally introduced by the
substitution in (\ref{8a}) \be 2qx^i\to\Phi^{(i)}(x)=\frac{2\pi}a
x^i+\phi^{(i)}(x)/f\,,\label{small}\ee where the three scalar
fields $\Phi^{(i)}$  satisfy \be
\langle\Phi^{(i)}\rangle_0=\frac{2\pi}ax_i\label{vacuum}\ee
whereas for the phonon fields one has \be
\langle\phi^{(i)}(x)\rangle_0=0\ .\ee In (\ref{small}) $f$ is the
decay constant of the phonon with dimension of an energy. We have
therefore three fluctuating fields $\phi^{(i)}_{k_1\,k_2\,k_3}$
for any elementary cube defined by discrete coordinates \be
x_{k_1}=\frac{k_1\pi}q~,~~y_{k_2}=
\frac{k_2\pi}q~,~~z_{k_3}=\frac{k_3\pi}q~,~~\ee
 i.e. \be
\phi^{(i)}_{k_1\,k_2\,k_3}\equiv\phi^{(i)}
(t,x_{k_1},y_{k_2},z_{k_3})\ .\label{83}\ee The interaction term
with the NGB fields  will be therefore given by\bea
S_{int}&=&-\int dt\left( \frac{\pi}{q}\right)^3
\sum_{k_1=-\infty}^{+\infty}\sum_{k_2=-\infty}^{+\infty}
\sum_{k_3=-\infty}^{+\infty}\, \sum_{\vec v}\sum_{m=1}^8\Delta
\exp\{i\,\varphi^{(m)}_{k_1\,k_2\,k_3}/f\}\cr
&&\cr&\times&\epsilon_{ij} \epsilon^{\alpha\beta
3}\psi^T_{i,\alpha,\vec v}\,C\,\psi_{j,\beta,-\vec v}-\,(L\to R)\
+\ h.c. \label{external0}\eea where \be
\varphi^{(m)}_{k_1\,k_2\,k_3}=\sum_{i=1}^3\epsilon_i^{(m)}
\phi^{(i)}_{k_1\,k_2\,k_3}\ee and the eight vectors $
\vec\epsilon^{(m)}$ are given by \be
(\epsilon_i^{(m)})\equiv\vec\epsilon^{\ (m)}\equiv \sqrt 3 \,\vec
n_m . \ee Notice that  translating the fields in momentum space
one would have to consider Fourier series with respect to the
discrete coordinates. This introduces in the theory a
quasi-momentum $\vec p$ through the combination
$\exp(i(p_xx_{k_1}+p_yy_{k_2}+p_zz_{k_3})$. Therefore one has to
restrict $\vec p$ to the first Brillouin zone $-q\le p_x\le q$,
$-q\le p_y\le q$, $-q\le p_z\le q$.

 The complete effective action for the NGB fields $\phi^{(i)}$
will be of the form \be S=\int dt\left( \frac{\pi}{q}\right)^3
\sum_{k_1=-\infty}^{+\infty}\sum_{k_2=-\infty}^{+\infty}
\sum_{k_3=-\infty}^{+\infty}{\cal
L}(\phi^{(i)}(t,k_1\pi/q,k_2\pi/q,k_3\pi/q))\ .\ee In the action
bilinear terms of the type
$\phi^{(i)}_{k_1\,k_2\,k_3}\phi^{(i^\prime)}_{k^\prime_1\,k^\prime_2\,k^\prime_3
}$ with $\{k_1\,k_2\,k_3\}\neq
{k^\prime_1\,k^\prime_2\,k^\prime_3}$ may arise. In the continuum
limit this terms would correspond to partial derivatives with
respect to the three space directions.

All these considerations become superfluous  when one realizes
that we are actually interested in an effective description of the
fields $\phi^{(i)}_{k_1,k_2,k_3}$ in the low energy limit, i.e.
for wavelengths much longer than the lattice spacing $\sim 1/q$.
In this limit the fields $\phi^{(i)}_{k_1,k_2,k_3}$ vary almost
continuously and can be imagined as continuous functions of three
space variables $x$, $y$ and $z$. Therefore we shall use in the
sequel the continuous notation $\phi^{(i)}(t,\vec r)$.

 The
coupling of the quark fields to the NGB fields generated by the
condensate will be written as
 \be
 \Delta\psi^TC\psi\sum_{\epsilon_i=\pm}\exp\left\{
i(\epsilon_1\Phi^{(1)}+\epsilon_2\Phi^{(2)}+\epsilon_3\Phi^{(3)})\right\}
 \label{coupling}\ee
making the theory invariant under translations and rotations.
These invariances are broken spontaneously in the vacuum defined
by eq. (\ref{vacuum}).

Integrating out the Fermi degrees of freedom one gets an effective
lagrangian containing only the light degrees of freedom, i.e. the
gluons of the unbroken subgroup $SU(2)_c$ and the phonons. We
will discuss gluons in Section \ref{gluecube} and we consider
here only the phonon effective lagrangian.

 The effective
lagrangian for the fields $\Phi^{(i)}$ has to enjoy the following
symmetries: rotational and translational invariance;
 $O_h$ symmetry on the fields $\Phi^{(i)}$.
The latter requirement follows from the invariance of the coupling
(\ref{coupling}) under the group $O_h$ acting upon $\Phi^{(i)}$.
The phonon fields $\phi^{(i)}(x)$ and the coordinates $x^i$ must
transform under the diagonal discrete group obtained from the
direct product of the rotation group acting over the coordinates
and the $O_h$ group acting over $\Phi^{(i)}(x)$. This is indeed
the symmetry left after the breaking of translational and
rotational invariance.

In order to build up the effective lagrangian, we start noticing
that if $X_i$ are quantities transforming as the same
representation of the $\Phi^{(i)}$'s under $O_h$ (a 3-dimensional
representation), then all the invariant expressions can be
obtained by the following three basic invariant expressions \be
I_2(X_i)=X_1^2+X_2^2+X_3^2,~~I_4(X_i)=X_1^2 X_2^2 +X_2^2 X_3^2+
X_3^2 X_1^2,~~ I_6(X_i)=X_1^2 X_2^2 X_3^2, \label{8}\ee which is a
general property of the symmetric functions of three variables.
The invariance is easily checked by noticing that these
expressions are invariant under the three elementary rotations
$R_i$  and the inversion $I$ (see eq. (\ref{basic})). Quantities
transforming as $\Phi^{(i)}$ are \be \dot\Phi^{(i)}\equiv
\frac{d\Phi^{(i)}}{dt}~,~~~ \vec\nabla\Phi^{(i)}\ .\ee We proceed
now as in ref. \cite{casa}, by noticing that the most general
low-energy effective lagrangian, invariant under rotations,
translations and $O_h$ is given by \be {\cal L}=\frac {f^2}
2\sum_{i=1,2,3}({\dot\Phi}^{(i)})^2+ {\cal L}_{\rm
s}(I_2(\vec\nabla\Phi^{(i)}),
I_4(\vec\nabla\Phi^{(i)}),I_6(\vec\nabla\Phi^{(i)}))\ .\ee The
reason for allowing all the possible spatial gradients in this
expansion is that the vacuum expectation value of
$\vec\nabla\Phi^{(i)}$ is proportional to $q$ and it is not in
general small. By following ref. \cite{crystal} one gets
 the effective lagrangian for the fields
$\phi^{(i)}$ at the second order in the derivatives as follows \be
{\cal L}=\frac 1 2\sum_{i=1,2,3}({\dot\phi}^{(i)})^2-\frac a 2
\sum_{i=1,2,3}|\vec\nabla\phi^{(i)}|^2- \frac b 2
\sum_{i=1,2,3}(\de_i\phi^{(i)})^2-
c\sum_{i<j=1,2,3}\de_i\phi^{(i)}\de_j\phi^{(j)}\,.\label{effective}\ee
The effective theory depends on three arbitrary parameters, a
result that follows also from the general theory of the
propagation of elastic waves in crystals \cite{LL}.
\section{Parameters of the effective lagrangian for the
phonon fields\label{sec6}} In this Section we will  compute the
parameters $a,b,c$ appearing in (\ref{effective}). We use the
gradient expansion. First we introduce external fields with the
same quantum numbers of the NGB's and then we perform a derivative
expansion of the generating functional. This gives rise to the
effective action for the NGB. At the lowest order one has to
consider the diagrams in Fig. 2, i.e. the self-energy, Fig. 2a,
and the tadpole, Fig. 2b.
\begin{figure}
\epsfxsize=8truecm \centerline{\epsffile{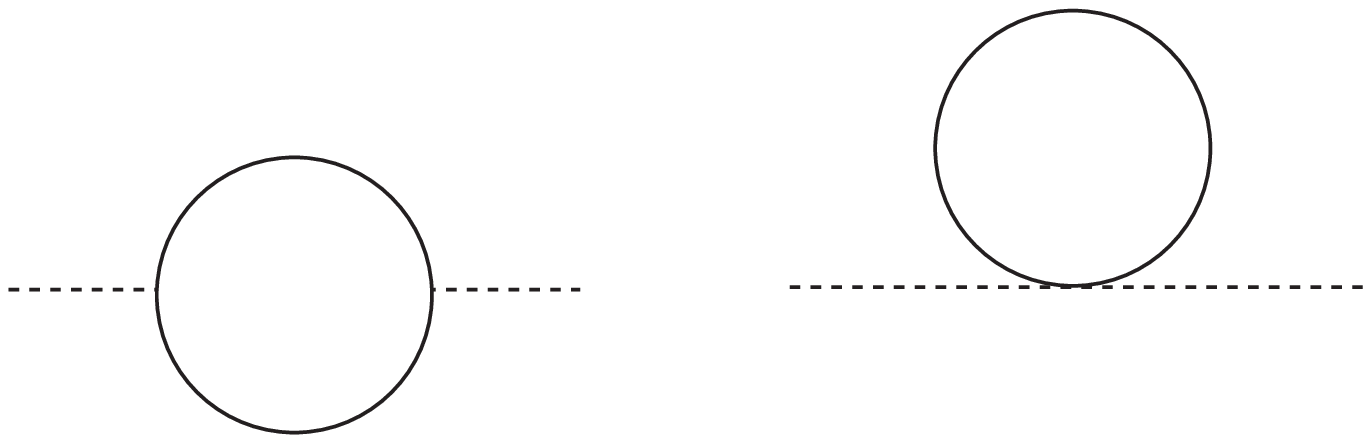}} \noindent
\hskip 4.5cm a) \hskip 4cm b) \par\noindent \vskip0.5cm Fig. 2.
Self-energy (a) and tadpole (b) diagrams.
\end{figure}
The diagrams can be computed observing that, by expanding
(\ref{coupling}) in the phonon fields, one has couplings of two
quarks with one and two phonons. Let us put
\be\varphi^{(m)}(t,\vec r)=\sum_{i=1}^3\epsilon_i^{(m)}
\phi^{(i)}(t,\vec r) \ .\ee

 At the first order in the fields
one gets the coupling with one NGB
 \bea
 {\cal L}_{\phi\psi\psi}&=&- \sum_{\vec
v}\sum_{m=1}^8\frac{\pi\Delta}{R}\,\delta_R[h(\vec v\cdot\vec
n_m)]\frac {(i\,\varphi^{(m)})}f
 \epsilon_{ij}\epsilon^{\alpha\beta 3}\psi_{\vec
v;\,i\alpha}\,C\,\psi_{-\vec v;\,j\beta} \cr  &-&(L\to R)\ +\
h.c.\label{Trilineare}\eea and the coupling of two quarks and two
NGB's \bea
 {\cal L}_{\phi\phi\psi\psi}&=&\sum_{\vec
v}\sum_{m=1}^8\frac{\pi\Delta}{R}\,\delta_R[h(\vec v\cdot\vec
n_m)]\frac{(\varphi^{(m)})^2}{2f^2}\
\epsilon_{ij}\epsilon^{\alpha\beta 3}\psi_{\vec
v;\,i\alpha}\,C\,\psi_{-\vec v;\,j\beta} \cr  &-&(L\to R)\ +\ h.c.
 \label{quadrilineare}\eea

In the basis of the $\chi$ fields:
 \be {\mathcal L}_3\,+\,{\mathcal L}_4=
 \sum_{\vec v}\sum_{A=0}^3
\tilde\chi^{A\,\dag}\,   \left(
\begin{array}{cc}
 0 & g_3^\dag\,+\,g_4^\dag
\\
  g_3\,+\,g_4& 0\end{array}\right)
    \,\tilde\chi^B\ ,\label{vertex}\ee
Here \bea g_3&=&\sum_{m=1}^8\frac{\pi\Delta}{R}\,\delta_R[h(\vec
v\cdot\vec n_m)]\frac{i\varphi^{(m)}}f\left(\begin{array}{cccc}
   1& 0 & 0 & 0 \\
  0 & -1 &0 & 0 \\
  0 & 0&-1& 0 \\
  0 & 0 & 0 & -1
\end{array}\right)_{AB}\ ,\cr&&\cr
g_4&=&\,-\sum_{m=1}^8\frac{\pi\Delta}{R}\,\delta_R[h(\vec
v\cdot\vec
n_m)]\frac{(\varphi^{(m)})^2}{2f^2}\left(\begin{array}{cccc}
   1& 0 & 0 & 0 \\
  0 & -1 &0 & 0 \\
  0 & 0&-1& 0 \\
  0 & 0 & 0 & -1
\end{array}\right)_{AB}\eea

To  perform the calculation one employs the propagator given in
Eq. (\ref{propagatore}) with $\Delta_{eff}$ given in
(\ref{deltaeff3}) and the interaction vertices in (\ref{vertex}).
The result of the calculation of the two diagrams at the second
order in the momentum expansion is as follows: \bea
 {\cal L}_{eff}(p)_{s.e.}&=&i\,
 \frac{4\times 4\,\mu^2}{16\pi^3f^2} \sum_{\vec
v}\sum_{m,k=1}^8\frac 1
2\left(\frac{\pi\Delta}{R}\right)^2\,\delta_R[h(\vec v\cdot\vec
n_m)](i\,\varphi^{(m)})\cr&&\cr&\times&\delta_R[h(\vec v\cdot\vec
n_k)](i\,\varphi^{(k)})\int
\frac{d^2\ell}{D(\ell)D(\ell+p)}\cr&&\cr&\times&
\Big[-2\Delta_{eff}^2+V\cdot\ell\,\tilde V\cdot(\ell+p)+\tilde
V\cdot\ell\, V\cdot(\ell+p) \Big]\ ,
 \cr
 &&
 \cr
 &&
 \cr
{\cal L}_{eff}(p)_{tad}
 &=&
 i\,\frac{4\times 4\mu^2}{16\pi^3f^2}
  \sum_{\vec v} \,\sum_{m=1}^8 \int  \frac
{d^2\ell}{D(\ell)}\frac{\pi\Delta\Delta_{eff}}{R}\,\delta_R[h(\vec
v\cdot\vec n_m)](\varphi^{(m)})^2\nonumber
 \eea
  where
  \be
D(\ell)=\ell_0^2-\ell_\parallel^2-\Delta^2_{eff}+i\epsilon\hskip0.3cm
, \ee
 One can easily control that the Goldstone theorem is
satisfied and the phonons are massless. As a matter of fact one
has \bea{\cal L}_{mass}&=&{\cal L}_{eff}(0)_{s.e.}\,+\,{\cal
L}_{eff}(0)_{tad}\, =i\,\frac{4\times
4\mu^2}{16\pi^3f^2}\,\frac{\pi\Delta}{R}\times\cr&&\cr&\times&
  \sum_{\vec v}\int  \frac
{d^2\ell}{D(\ell)}\Big[-\sum_{m,k=1}^8\frac{\pi\Delta}{R}\,\delta_R[h(\vec
v\cdot\vec n_m)]\varphi^{(m)}\delta_R[h(\vec v\cdot\vec
n_k)]\varphi^{(k)}\cr&+&\Delta_{eff}\sum_{m=1}^8 \delta_R[h(\vec
v\cdot\vec n_m)](\varphi^{(m)})^2\Big]\ .\label{45}\eea

Both terms in the r.h.s. of eq. (\ref{45}) present  a double sum:
$m,k=1,...,8$ (see eq. (\ref{deltaeff3}) for $\Delta_{eff}$); we
have seen at the end of section \ref{hdet} that the eight rings
that define the pairing region are non overlapping. Therefore only
the terms with $m=k$ survive and one immediately verifies the
validity of the Goldstone's theorem. i.e. the vanishing of
(\ref{45}). Notice that in our approximation the masses of the
Goldstone bosons vanish only if the pairing regions are not
overlapping, signaling that when the regions overlap we are not at
the minimum of the free-energy (see \cite{bowers}).

 At the second order in the momentum
expansion one has \bea {\cal L}_{eff}(p)&=&i\,
 \frac{4\times 4\,\mu^2}{16\pi^3f^2} \sum_{\vec
v}\sum_{m,k=1}^8\frac 1
2\left(\frac{\pi\Delta}{R}\right)^2\,\delta_R[h(\vec v\cdot\vec
n_m)](i\,\varphi^{(m)})\cr&&\cr&\times&\delta_R[h(\vec v\cdot\vec
n_k)](i\,\varphi^{(k)})\int d^2\ell\,\frac{2\Delta_{eff}^2\,
V\cdot p\,\tilde V\cdot p}{[D(\ell)]^3}\ . \eea Using the result
\be \int
\frac{d^2\ell}{[D(\ell)]^3}\,=\,-\,\frac{i\pi}{2\Delta_{eff}^4}\
,\ee and the absence of off-diagonal terms in the double sum, we
get the effective lagrangian in the form
 \bea
 {\cal L}_{eff}(p)&=&-\,\frac{\mu^2}{2\pi^2 f^2}\sum_{\vec v}
\left(\frac{\pi\Delta}{R}\right)^2\,\sum_{k=1}^8\frac{(\delta_R[h(\vec
v\cdot\vec n_k)])^2}{\Delta_{eff}^2}\cr&\times&\left( V\cdot
p\right)\varphi^{(k)}\left(\tilde V\cdot p\right)\varphi^{(k)} \
.\label{seinove} \eea To perform the calculation one can exploit
the large value found per $R$ , i.e. eq. (\ref{RT}). Since for
$R\to\infty$ the  $\delta_R$ function becomes the Dirac delta, we
take the $R\to\infty$ limit and we handle the $\delta_R$ function
according to the Fermi trick in the Golden Rule; in the numerator
we   substitute one $\delta_R$ function with the Dirac delta while
for the other one we take
 \be
\frac{\pi\delta_R[h(x)]} R\to\frac{\pi\delta_R(0)}R\to \,1 .\ee
We observe that this limit is justified by the large value found
for $R$, see eq. (\ref{RT}). The  sum over $k$ in (\ref{seinove})
gives \bea &&\sum_{k=1}^8\delta_R[h(\vec v\cdot\vec
n_k)]\varphi^{(k)}\delta_R[h(\vec v\cdot\vec
n_k)]\varphi^{(k)}\to\,\frac{R}{\pi}\, \sum_{k=1}^8\delta[h(\vec
v\cdot\vec n_k)]\left(\varphi^{(k)}\right)^2=\cr &&=\frac
R\pi\sum_{k=1}^8\,\delta\left[1-\frac{\delta\mu}{q\vec v\cdot\vec
n_k}\right]\left(\varphi^{(k)}\right)^2=\frac{
R^2}{\pi^2}k_R\sum_{k=1}^8\delta\left[\vec v\cdot\vec
n_k-\frac{\delta\mu}{q}\right]\left(\varphi^{(k)}\right)^2 \eea
with  \be k_R=\frac{\pi|\delta\mu|}{qR} \ .\ee On the other hand
$\Delta_{eff}^2$ in the denominator of (\ref{seinove}) gives \be
\Delta_{eff}^2 \ \to \ \Delta^2 \ .\ee Therefore one gets
 \be{\cal
L}_{eff}(p)\,=\,-\,\frac{\mu^2k_R}{2\pi^2f^2}\sum_{i,j=1}^3\sum_{k=1}^8\sum_{\vec
v } \delta\left\{\vec v\cdot\vec n_k-\frac{\delta \mu}{q}\right\}
V_\mu\tilde V_\nu \epsilon^{(k)}_i\epsilon^{(k)}_j
p_\mu\phi^{(i)}p_\nu\phi^{(j)}\ . \ee The integration over the
Fermi velocities require special attention. We use the result
 \be
 \sum_{k=1}^8
 \epsilon^{(k)}_i\epsilon^{(k)}_j \,=\,8\delta_{ij}\,;\label{74}\ee
 this fixes the constant multiplying the time derivative term in
 the effective lagrangian at the value (taking into account
 (\ref{sumvel}))
 \be \frac{8\mu^2k_R}{4\pi^2 f^2}\ .
 \ee Therefore
  one obtains canonical normalization for the kinetic term
provided \be
 f^2= \frac{8\mu^2k_R}{2\pi^2}
  \ .
\ee To compute the parameters $a,\,b,\,c$ of the effective
lagrangian (\ref{effective}) we need to evaluate \be \beta_{\ell
m}^{ij}= \sum_{k=1}^8
 \epsilon^{(k)}_i\epsilon^{(k)}_j\int\frac{d\vec v }{8\pi}v_\ell\,
 v_m\,
\delta\left\{\vec v\cdot\vec n_k-\frac{\delta \mu}{q}\right\}\
.\label{57} \ee We find the result (see Appendix 2) \be
\beta_{\ell m}^{ij}=\frac 2 3\delta_{lm}\delta_{ij}+
\frac{3\cos^2\theta_q-1}{3}\sum_{r=1}^3\rho^r_{lm} \rho^r_{ij}
\ee where \be
\rho^r_{ij}=\delta_{is}\delta_{jt}+\delta_{it}\delta_{js},~~~~r,s,t~{\rm
in~ cyclic~ order}\,.\ee From this equation we get \bea &&{\cal
L}_{eff}(p)=\frac 1 2\left( {p^0}^2{\phi^{(i)}}^2-\frac 1
8\beta^{ij}_{lm}p^l p^m\phi^{(i)}\phi^{(j)}\right)=\nn\\&=&\frac
1 2 \Big({p^0}^2{\phi^{(i)}}^2-\frac {|\vec p|^2}{12}
{\phi^{(i)}}^2-\frac{3\cos^2\theta_q-1}6\sum_{i<j=1,2,3}p^i\phi^{(i)}p^j\phi^{(j
)}\Big)\ . \eea

Comparing with eq. (\ref{effective}) we finally obtain \be
a=\frac{1}{12} \ ,~~~~~ b=0\ ,~~~~~
 c=\frac{3\cos^2\theta_q-1}{12}\ .\ee One can note that these
 values for the parameters of the phonon effective lagrangian
 satisfy the positivity constraints found in \cite{crystal}.

In \cite{crystal} one can find the general discussion of the
dispersion relation for arbitrary values of the parameters $a$,
$b$ and $c$. In the present paper we give the expression for the
eigenvalues of $E^2$, according to the direction of propagation of
the momentum $\vec p$, in  table 1.
\begin{center}
\begin{tabular}{||c|c|c||}
\hline\hline &&\\ Momentum & Eigenvalues & Eigenvectors\\&&\\
\hline\hline
 &&\\ \fbox{$C_4$}~~~~~~~~~~~~~~~~~~~~~~~~~~~~~~~~~~~~
 & $1/12$& $\phi^{(i)}$ \\
 $\vec p = p\, {\vec e\,}^{(i)} $ & $ 1/12$ &
 $\phi^{(j)}$, $\phi^{(k)}$ ($i\not=j\not=k$)\\
 &&\\ \hline\hline&&\\ \fbox{$C_3$}~~~~~~~~~~~~~~~~~~~~~~~~~~~~~~~~~~~~
 & $(1+3\sin^2\theta_q)/6$&
 $-\phi^{(1)}+x\phi^{(2)}+(1-x)\phi^{(3)}$ \\
 $\vec p = p\,({\vec
e\,}^{(1)}+{\vec e\,}^{(2)}+{\vec e\,}^{(3)}) $ &
$(1+6\cos^2\theta_q) /36$ &
 $\phi^{(1)}+\phi^{(2)}+\phi^{(3)}$\\
&&\\
 \hline &&\\
& $(1+3\sin^2\theta_q)/6$& $\phi^{(1)}+x\phi^{(2)}+(1-x)\phi^{(3)}$ \\
 $\vec p = p\,(-{\vec
e\,}^{(1)}+{\vec e\,}^{(2)}+{\vec e\,}^{(3)}) $ &
$(1+6\cos^2\theta_q) /36$
& $-\phi^{(1)}+\phi^{(2)}+\phi^{(3)}$\\&&\\
 \hline &&\\
& $(1+3\sin^2\theta_q)/6$& $\phi^{(1)}+x\phi^{(2)}-(1-x)\phi^{(3)}$ \\
 $\vec p = p\,({\vec
e\,}^{(1)}-{\vec e\,}^{(2)}+{\vec e\,}^{(3)}) $ &
$(1+6\cos^2\theta_q) /36$
& $\phi^{(1)}-\phi^{(2)}+\phi^{(3)}$\\&&\\
 \hline &&\\
& $(1+3\sin^2\theta_q)/6$& $\phi^{(1)}+x\phi^{(2)}+(1+x)\phi^{(3)}$ \\
 $\vec p = p\,({\vec
e\,}^{(1)}+{\vec e\,}^{(2)}-{\vec e\,}^{(3)}) $ &
$(1+6\cos^2\theta_q) /36$
& $\phi^{(1)}+\phi^{(2)}-\phi^{(3)}$\\&&\\
 \hline\hline &&\\ \fbox{$C_2$}~~~~~~~~~~~~~~~~~~~~~~~~~~~~~~~~~~~~
 & $1/12$& $\phi^{(k)}$ \\
 $\vec p = p\,({\vec
e\,}^{(i)}+{\vec e\,}^{(j)}) $ & $\sin^2\theta_q /8$ &
 $-\phi^{(i)}+\phi^{(j)}$\\
 $i<j$&$(1+3\cos^2\theta_q)/24$ &$\phi^{(i)}+\phi^{(j)}$\\&&\\
 \hline &&\\
&$1/12$& $\phi^{(k)}$ \\
 $\vec p = p\,({\vec
e\,}^{(i)}-{\vec e\,}^{(j)}) $ & $\sin^2\theta_q /8$ &
 $\phi^{(i)}+\phi^{(j)}$\\
 $i<j$&$(1+3\cos^2\theta_q)/24$ &$-\phi^{(i)}+\phi^{(j)}$\\&&\\
 \hline\hline
\end{tabular}
\end{center}
{\bf Table 1} - {\it Eigenvalues and eigenvectors of
$\frac{E^2}{|\vec p|^2}$; $\vec p$ along the symmetry axes.}

\section{Gluon dynamics in the the LOFF
phase: single plane wave structure }In this Section and in the
subsequent one we wish to derive the effective lagragian for the
gluons of the unbroken $SU(2)_c$ subgroup of the two-flavor LOFF
phase.  To start with we review the results obtained for the
gluon effective lagrangian at one-loop in the $2SC$ isotropic
phase \cite{abr}. They were obtained  in \cite{son}; in
\cite{Casalbuoni:2001ha} the results of \cite{son} were obtained
by the method of the High Density Effective Theory  and extended
to the three-gluon and four-gluon vertices. In the $2SC$ phase
one considers massless quarks of two different flavors at the
same chemical potential $\mu$ and at low temperature. Color
superconductivity breaks the color group $\dd SU(3)_c$ down to a
$\dd SU(2)_c$
 subgroup and the quarks with nontrivial $\dd SU(2)_c$ charges
acquire a gap $\dd \Delta_{0}$. Therefore only the three gluons
relative to the generators of the unbroken subgroup remain
massless.

The lagrangian including one loop corrections for the $\dd
SU(2)_c$ gluons can be written as \cite{son},
\cite{Casalbuoni:2001ha} \be {\mathcal L}\ =\ \frac{1}{2} (E_i^a
E_i^a - B_i^a B_i^a) + \frac{k}{2} E_i^a E_i^a ~, \label{lag} \ee
with $a=1,2,3$ color indices and \be k= \frac{g^2 \mu^2}{18 \pi^2
\Delta_{0}^2} \label{glue} \ .\ee Eqns. (\ref{lag}) and
(\ref{glue}) mean that, at one loop, the medium has a {\it
dielectric constant} \be \epsilon = k +1\label{iso} \ee and a {\it
magnetic permeability} $\dd \lambda = 1$, while the gluon speed in
the medium is \be v=\frac{1}{\sqrt{\epsilon \lambda}} \propto
\frac{ \Delta_{0}}{g \mu} ~. \ee  As noted in \cite{son}, it is
possible to put (\ref{lag}) in a gauge invariant form \be
{\mathcal L}\ =\ - \frac{1}{4} F^{\mu \nu}_a F^{a}_{\mu \nu} ~~~~
(a=1,2,3) ~, \label{36} \ee provided the following rescaling is
used \be A_0^a \rightarrow A_0^{a \prime} = k^{3/4} A_0^a ~,
\label{A0} \ee \be A_i^a \rightarrow A_i^{a \prime} =  k^{1/4}
A_i^a ~, \label{Ai} \ee \be x_0 \rightarrow x_0^{\prime}  =
k^{-1/2} x_0 ~, \label{x0} \ee \be g \rightarrow g^{\prime}  =
k^{-1/4} g ~. \label{g} \ee

Let us now consider the LOFF non-isotropic phase. To begin with we
assume the crystal structure given by a plane wave i.e. $
\Delta(\vec x)$ given by (\ref{deltaeff}). The difference with the
isotropic case consists, in the present case, in the substitution
\be \Delta\to\Delta_{eff}=\frac{\Delta\pi} R\delta_R[h(\vec
v\cdot\vec n)] \label{deltaeff1} ~,\ee where $\vec n=\vec q/q$ and
$h(x)$ given by (\ref{hdix}).

The effective action allows the evaluation of the one loop
diagrams with two external gluon lines and internal quark lines
similar to those in fig. 2. Their contribution to the polarization
tensor is written down in Appendix 3.

 Let us now write\be
\Pi^{\mu \nu}_{a b}(p)\ =\ \Pi^{\mu \nu}_{a b}(0) + \delta
\Pi^{\mu \nu}_{a b}(p) \label{pi}\,.\ee As discussed in Appendix
3 we find a vanishing Meissner mass \be \Pi_{ab}^{ij}(0)=0\ee and
a non-vanishing Debye screening mass \be m_D = \frac{g \mu}{\pi}
\sqrt{ 1 + \frac{\cos \theta_a-\cos\theta_b}{2} }
\label{120}\,,\ee where  $\cos\theta_a$ and $\cos\theta_b$
($-1\le\cos\theta_a\le\cos\theta_b\le 1$) are the solutions of
the equation\be |h(\cos\theta)|=\frac \pi{2R}\,.\label{106}\ee
where $h(\cos\theta)$ is defined in eq. (\ref{29}).

It is interesting to note that, in absence of LOFF pairing, i.e.
if the pairing region vanishes and $\cos\theta_a=\cos\theta_b$,
one gets for the Meissner and Debye mass the same results
obtained  in the hard dense loop approximation for the model with
two flavors in the normal (not superconducting) state, see e.g.
\cite{r2}. On the other hand if the blocking region vanishes
($\cos\theta_a=-1$, $\cos\theta_b=+1$) one is in the 2SC phase and
eq. (\ref{120}) gives $m_D=0$ as expected.

Next consider $\delta \Pi^{\mu \nu}_{a b}(p)$. The contribution of
the blocking region  vanishes as discussed in Appendix 3.
Therefore the only non vanishing contribution to $\delta \Pi^{\mu
\nu}_{a b}(p)$ comes from the pairing region. It could be
 computed exactly, but the corresponding expressions are
 cumbersome and shed little light on the underlying physics.
 Therefore we prefer to work in the approximation of small
 momenta ($|p\,|\ll\Delta$). We find
\be -\,\delta \Pi^{\mu \nu}_{a b}(p)\ =\ \delta_{ab}\frac{\mu^2
g^2}{12 \pi^2} \sum_{\vec v;\,\, pairing}\frac{V^{\mu} V^{\nu}
(\tilde V \cdot p)^2 -
 \tilde V^{\mu} V^{\nu}( V \cdot p\ \tilde V \cdot p)
 +V\leftrightarrow\tilde V}{\Delta^2_{eff}}. \label{pi2}
\ee That is
 \bea -\delta\Pi^{0 0}_{ab}(p)&=&\delta_{ab} \frac{g^2 \mu^2}{3
\pi^2} \sum_{\vec v;\,pairing} \frac{v_i v_j}{\Delta_{eff}^2} p_i
p_j =\cr&=&\delta_{ab} \frac{g^2 \mu^2R^2}{3 \Delta^2 \pi^4}
\int_{pairing} \frac{d\cos\theta\,d\phi}{8 \pi} \frac{v_i
v_j}{\left(\delta_R[h(\cos\theta)]\right)^2}\, p_i p_j
\label{pi0pbis1} \eea where $\vec
v=(\sin\theta\cos\phi,\sin\theta\sin\phi,\cos\theta)$ and $\vec
n=\vec q/q$ along the $z-$axis. The integration domain is defined
by $\cos\theta_a<\cos\theta<\cos\theta_b$. We get therefore
 \be -\delta\Pi^{0 0}_{ab}(p)\ =\
\delta_{ab} k \left( f(R) p_\bot^2+
 g(R)p^2_\| \right)~,
\label{pi00} \ee where $k$ is given by (\ref{glue}) and \be f(R)
\, = \, \frac{3 }{4} \int_{pairing} d \cos \theta \left(1 -
\cos^2\theta\right)\ , \ee  \be g(R) \, = \, \frac{3}{2}
\int_{pairing} d \cos \theta\,\cos^2\theta \ee are functions of
the  parameter $R$ and are reported in fig. 3 ($q\approx1.2
|\delta\mu|$).

 It is interesting to note the anisotropy of the
polarization tensor exhibited by these results. One has always
$g>f$; for large $R$, and neglecting $\delta\mu/\mu$ corrections,
one finds approximately \be
\frac{g(R)}{f(R)}\,\to\,\frac{2}{\dd\left(\frac q
{\delta\mu}\right)^2-1}\ .
 \ee

\begin{figure}[htb] \epsfxsize=8truecm
\centerline{\epsfxsize=6truecm\epsffile{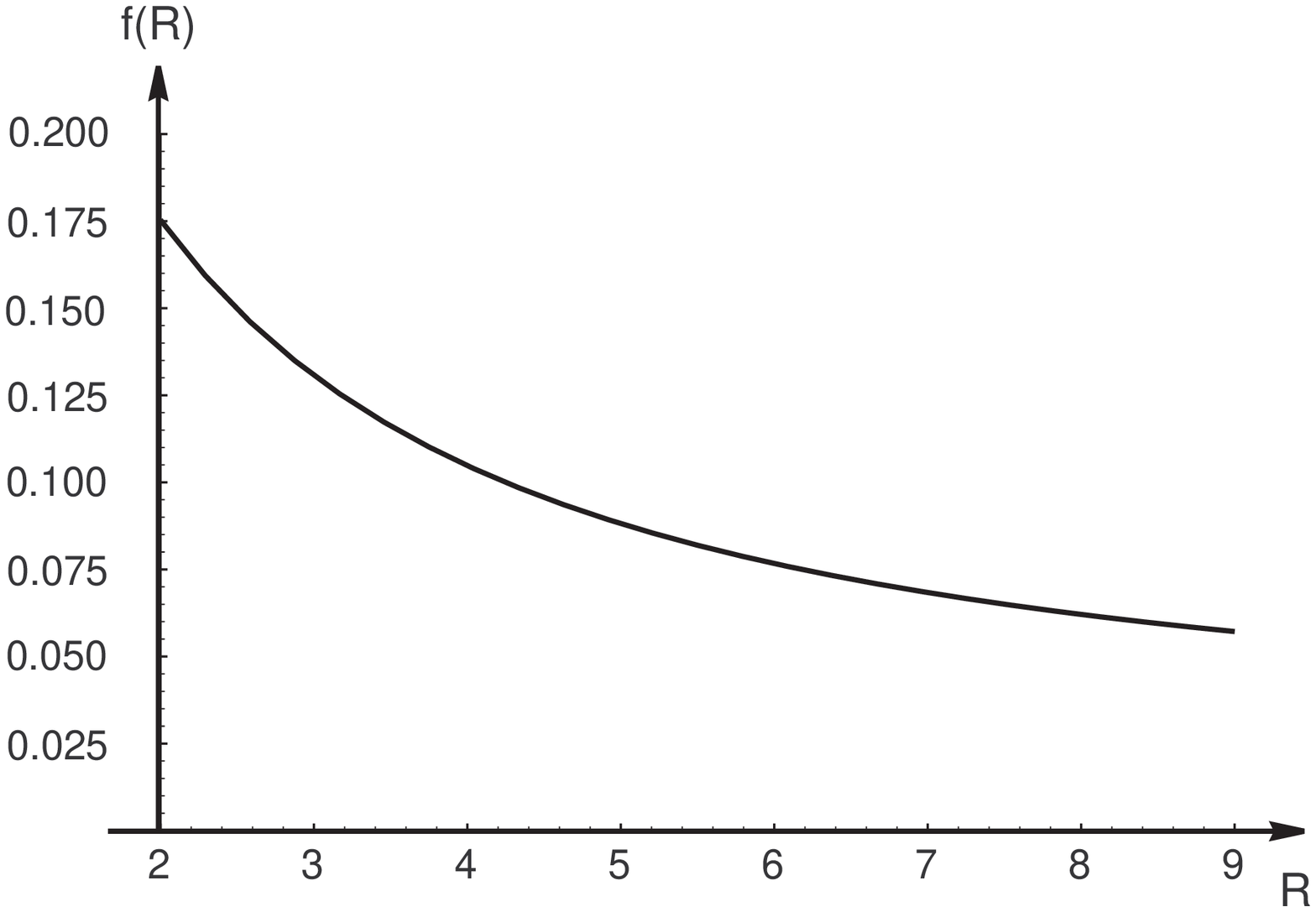}
\hskip1cm\epsfxsize=6truecm\epsffile{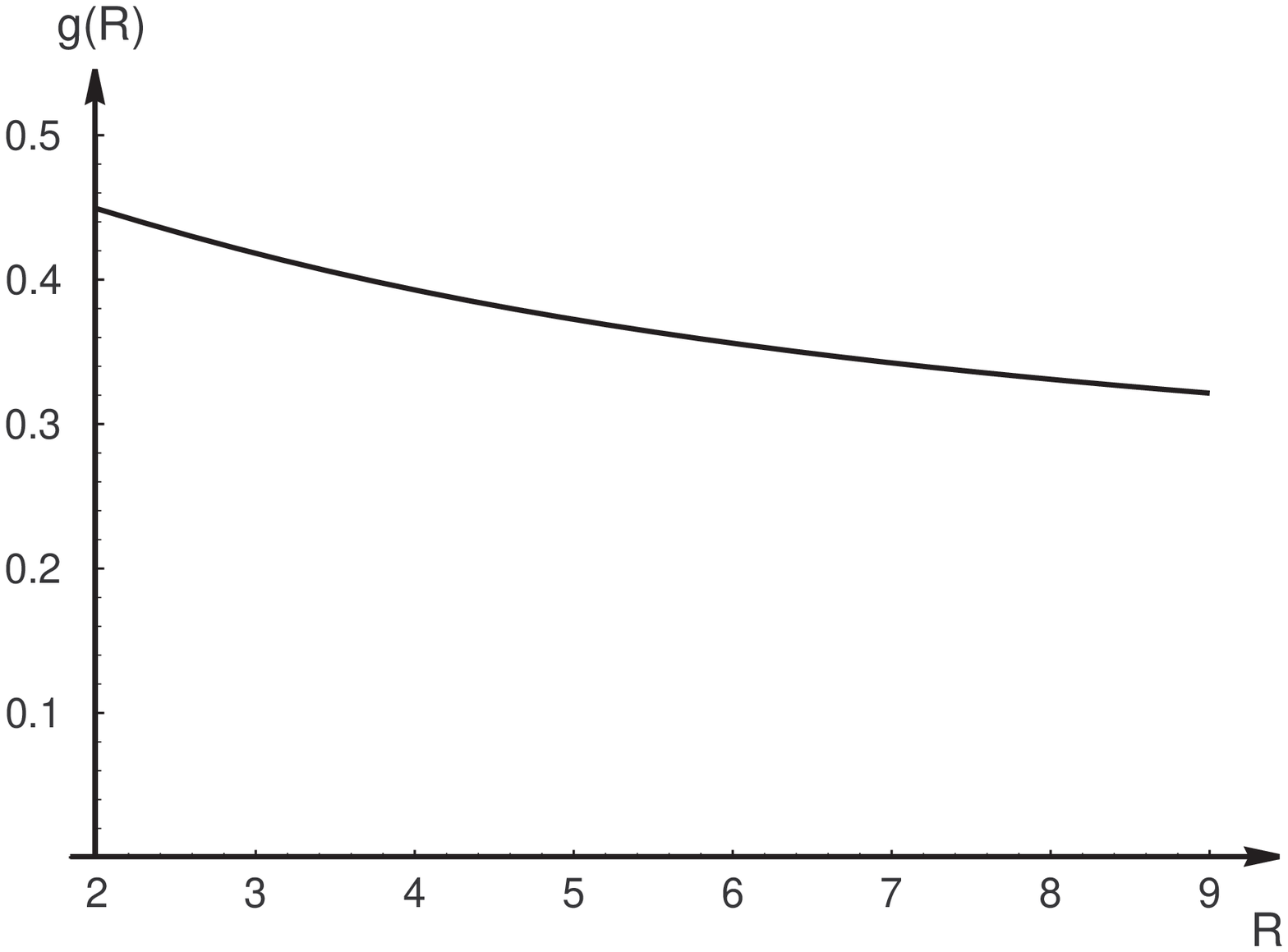}}
\vskip1cm\noindent \centerline{ Fig. 3 Plots of the functions
$f(R)$ and $g(R)$.}
\end{figure}

Let us finally write down the remaining components of the
polarization tensor. From (\ref{pi2}) we get \be
-\,\Pi^{ij}_{ab}(p)\ =\ \delta_{ab} \frac{g^2 \mu^2}{3 \pi^2}
\sum_{\vec v;\,\, pairing} \frac{v_i v_j}{ \Delta^2}  p_0^2 ~=\ k
\, p_0^2\ \Big( f(R) (\delta_{i1} \delta_{j1}
+\delta_{i2}\delta_{j2})+
 g(R) \delta_{i3} \delta_{j3} \Big)
\label{piijp} \ee and \be -\,\Pi^{0 i}_{ab}(p)\ =\ k \, p_0\,p^j\
\Big( f(R) (\delta_{i1} \delta_{j1} +\delta_{i2}\delta_{j2})+
 g(R) \delta_{i3} \delta_{j3} \Big) ~.
\label{pi0ip} \ee These results complete the analysis of the LOFF
model in the one plane wave approximation. From $\Pi^{\mu \nu}_{a
b}$ we get the dispersion law for the gluons at small momenta. The
lagrangian at one loop is\footnote{We do not include here the 3
and 4-gluon vertices that however can be handled as in
\cite{Casalbuoni:2001ha}, with the result that the local gauge
invariance of the one-loop lagrangian is satisfied.}\be {\mathcal
L}\ =\ -\frac{1}{4} F^{\mu \nu}_{a} F_{\mu \nu}^{a} - \frac{1}{2}
\Pi^{\mu \nu}_{a b} A_{ \mu}^a A_{ \nu}^b \label{su2lag2}  \ee
(sum over the repeated color indices $a,b=1,2,3$). Introducing the
fields $\dd E_i^a \equiv F_{0i}^a$ and $\dd B_i^a \equiv i
\varepsilon_{ijk} F_{jk}^a$, and using (\ref{pi00}), (\ref{piijp})
and (\ref{pi0ip}) we can rewrite the lagrangian (\ref{su2lag2}) as
follows \be {\mathcal L}\ =\ \frac{1}{2}
\left(\epsilon_{ij}\,E_i^a E_j^a  - B_i^a B_i^a\right)+\frac 1 2
m_D^2 A_a^0 A_a^0 \label{lag2} ~, \ee where\be
\epsilon_{ij}=\left( \matrix{1+kf(R) & 0& 0 \cr 0 & 1+kf(R) & 0
\cr 0 & 0 & 1+k g(R)}\right) ~. \ee This means that the medium
has a non-isotropic {\it dielectric tensor} $\dd \epsilon$ and a
{\it magnetic permeability} $\dd \lambda = 1$. These results have
been obtained taking the total momentum of the Cooper pairs along
the $z$ direction. Therefore we distinguish the dielectric
constant along the $z$ axis, which is \be \epsilon_{\parallel}
=1+ kg(R) ~, \ee and the dielectric constant in the plane
perpendicular to the $z$ axis \be \epsilon_{\perp} =1+ k f(R) ~.
\ee This means that the gluon speed in the medium depends on the
direction of propagation of the gluon; along the $z$ axis the
gluon velocity is \be v_{\parallel}\simeq\frac{1}{\sqrt{k g(R)}}
~, \ee while for gluons which propagate in the $x-y$ plane we
have \be v_{\perp}\simeq\frac{1}{\sqrt{k f(R)}} \ee and in the
limit of large $R$, and neglecting $\delta\mu/\mu$ corrections,
\be v_{\parallel}\to \frac 1{\sqrt{2}}\tan\theta_q\, v_\perp
\,.\ee with $\cos\theta_q$ defined in eq. (\ref{hdix}). The
anisotropy in the gluon velocities has its {\it pendant} in the
anisotropy of the phonon velocity \cite{effLOFF2}.
\section{Gluon dynamics in the LOFF
phase: cubic structure\label{gluecube}} The condensate in this
case is given by eq. (\ref{8a}), so that we will use the results
of sec. \ref{hdet} with $\Delta_{eff}$ given by (\ref{deltaeff3}).
 The  calculations  are similar to the previous
case and, similarly, the $SU(2)_c$ gluons have vanishing Meissner
mass and exhibit partial Debye screening.  However the dispersion
law of the gluons is different.

As a matter of fact we write  the  one loop lagrangian for the
$SU(2)_c$ gluons as\be {\mathcal L}\ =\ \frac{1}{2} (E_i^a E_i^a
 - B_i^a B_i^a) +\delta{\mathcal L}~, \label{multilag1} \ee
with \be\delta{\mathcal L}=-  \frac{1}{2} \Pi^{\mu \nu}_{a b} A_{
\mu}^a A_{ \nu}^b \label{su2lag1} ~.\ee In the approximation
$|p|\ll\Delta$, $\delta\Pi^{\mu \nu}_{a b}$ is again given by eq.
(\ref{pi2}), but now $\Delta_{eff}$ is given by
(\ref{deltaeff3}).  One gets \be \delta{\mathcal L}\equiv E^a_i
E^b_j \ \delta_{ab} \frac{g^2 \mu^2}{6 \pi^2} \int_{pairing}
\frac{d\cos\theta\,d\phi}{8\pi} \frac{v_i
v_j}{\Delta_{eff}^2}+A^a_0 A^b_0 \,\delta_{ab}\frac{g^2 \mu^2}{4
\pi^2} \int_{blocking} d\cos\theta ~. \label{pi0pbis} \ee
Evaluating the integrals one finds \be {\mathcal L}\ =\
\frac{1}{2} (\tilde\epsilon_{ij}\,E_i^a E_j^a  - B_i^a B_i^a)
+\frac 1 2 {M_D}^2A_0^a A_0^a~, \label{multilag2} \ee with the
tensor $\tilde\epsilon^{ij}$ given by \be \tilde\epsilon_{ij}
=\delta_{ij}\left[1\,+ \, k \,t(R)\right]
 ~.\ee and \be M_D=\frac{g\mu}\pi\sqrt{1+8\frac{\cos\theta_a-\cos\theta_b}2}\,.\ee where
 $\cos\theta_{a,b}$ are solutions of eq. (\ref{106}). It should also be noted that the
 values of the parameter $R$ for the cube and the plane wave can
 be different.

The isotropy of these results can be easily understood
 noticing that  $\delta{\mathcal L}$
is a symmetric function of
 its arguments. As a matter of fact it
  must be symmetric under the residual symmetry $O_h$
and therefore
 it must be built up from the three elementary symmetric functions
$I_2(X_i)$, $I_4(X_i)$ and $I_6(X_i)$, defined in eq. (\ref{8}),
with $X_i=E_i^a$. The invariance of $\delta{\mathcal L}$ follows
from the invariance of these functions under the three elementary
rotations $R_i$  and the inversion $I$. However it is clear from
(\ref{pi0pbis}) that $\delta{\mathcal L}$ is
 second-order in $E^i_a$ and therefore
 the only function that can be
 involved is $I_2$. In conclusion in the cubic
lattice one  necessarily gets \be
\tilde\epsilon_{ij}-\delta_{ij}=A\delta_{ij}\ .\ee

From eqs. (\ref{pi0pbis}) and (\ref{deltaeff3}) one has that the
function $t(R)$ is given by the formula \be t(R)\, \delta_{ij} \,
= \, \frac{3 R^2}{4 \pi^3} \int_{pairing} d\cos\theta d\phi
\frac{v_i v_j} {\left[ \sum_{k=1}^{8} \delta_R(h(\vec v \cdot \vec
n_k))\right]^2} ~. \ee $t (R) $ can be computed by observing that,
using the properties of the $\delta_R$ function \bea t(R)
\delta_{ij} \,& \approx& \, \frac{3 R^2}{4 \pi^3}\sum_{k=1}^8
\int_{pairing} d\cos\theta d\phi \frac{v_i v_j} {\left[
\delta_R(h(\vec v \cdot \vec n_k))\right]^2} =\cr&=&
\sum_{k=1}^8\left(f(R)\delta_{ij}+[g(R)-f(R)] n^{(k)}_i
n^{(k)}_j\right)\ .\eea Using (\ref{74}) we finally get \be t
(R)=\frac 8 3[2f(R)+g(R)] \ .\ee A plot of the function $t(R)$ is
in Fig. 4.
\begin{figure}[htb] \epsfxsize=8truecm
\centerline{
\epsfxsize=9truecm\epsffile{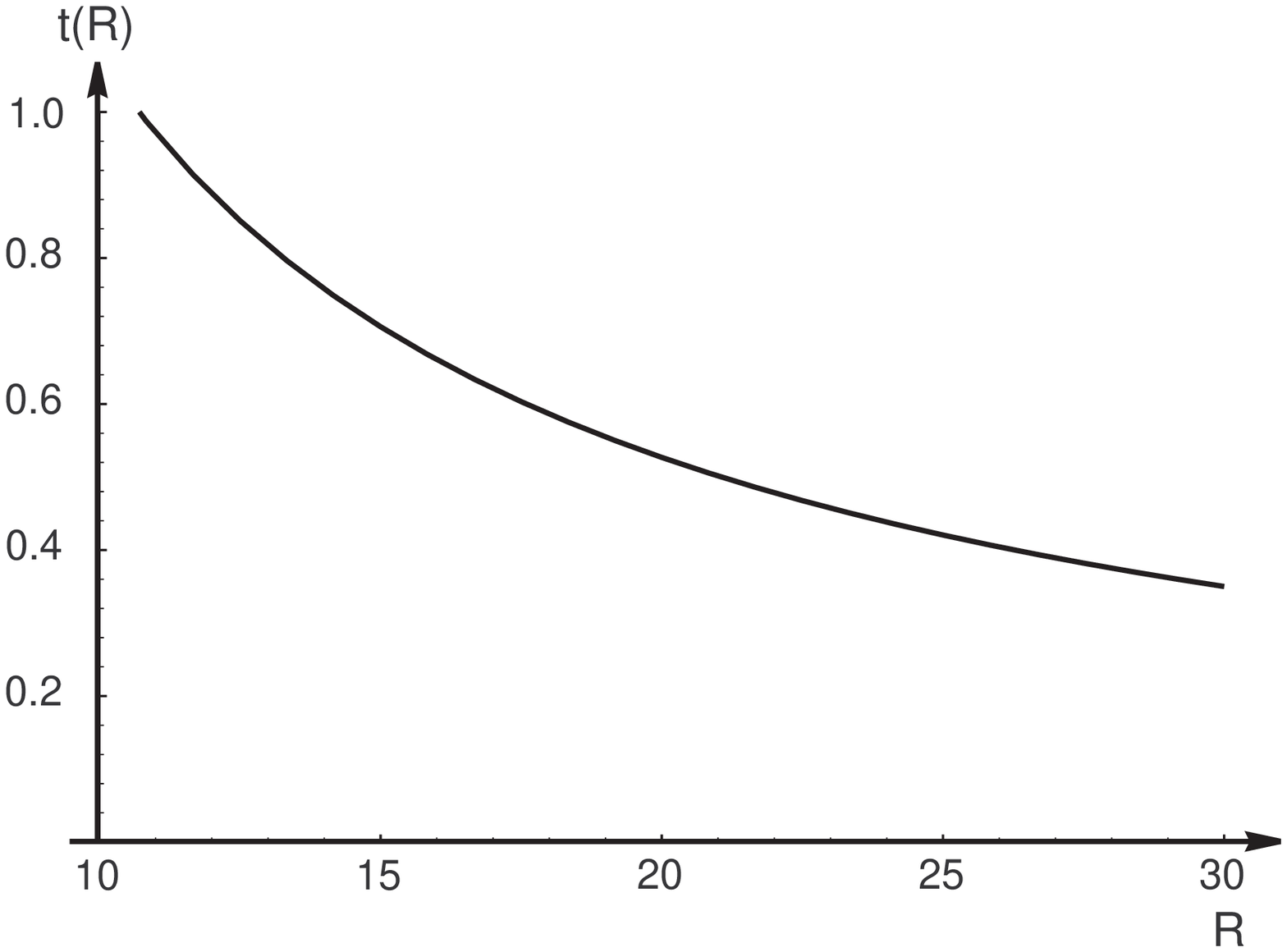}}\vskip1cm\noindent
\centerline{ Fig. 4 Plot of the function $t(R)$.}
\end{figure}

We finally note that, even if the crystalline structure is not
isotropic, in eq. (\ref{multilag2}) the $O(3)$ invariance is
restored. This happens because the tensor $\tilde\epsilon_{ij}$ is
proportional to the identity. Therefore the dielectric properties
of the medium will be isotropic and the velocity of propagation of
the gluons will be the same in all the directions. We stress that
these results are valid in the approximation of soft gluon
momenta.

\section{Conclusions}

We have used the High Density Effective Theory (HDET) formalism to
calculate the low energy properties of phonons and  gluons in the
LOFF phase of QCD.The HDET formalism is very useful to simplify
the dynamics as it allows direct elimination of the degrees of
freedom which are far from the Fermi surface. It bears useful
resemblances with the Heavy Quark Effective Theory of general use
when dealing with heavy hadrons.  We have first made the
modification for the case of a single plane wave condensate and
we have derived the quark propagators. After that we have gone to
the relevant case of the cubic strucure of ref. \cite{bowers}
corresponding to the sum of eight plane waves. We have added the
suitable term in the Lagrangian bearing the sum over the eight
plane waves of the cubic structure and derived for this more
complex case the quark propagators to be used in the calculations.

The three phonons correspond to the local deformations from the
fluctuations of the crystal as defined by the cubic condensate.
They are the Nambu-Goldstone bosons from the breaking of
translational symmetry. We have studied the low energy limit,
where wavelengths are much longer than lattice spacing, and built
up  an effective lagrangian for the phonons, which, at the second
order in the derivatives, depends on three parameters, in
agreement with the general theory of the elastic waves in crystals
\cite{LL}.

The parameters of the effective phonon lagrangian are calculated
by introducing external fields with the phonon quantum numbers and
performing the derivative expansion of the generating functional.
The phonons remain massless, as expected. The final results for
the parameters are remarkably simple, the only quantity appearing
in the result being the ratio of $\delta\mu$ to the common length
of the eight wave vectors appearing in the condensate.

We have also derived  the effective lagrangian of the gluons of
the unbroken $SU(2)_c$ subgroup. The gluon self energy diagrams
at one loop can be evaluated from the derived effective action.
The  Meissner mass vanishes as in the case of complete isotropy;
on the other hand there is a partial Debye screening originating
from the existence of blocking regions on the Fermi spheres. For
soft gluon momenta, in the case of a single plane wave, the
dielectric constant of the medium is non isotropic and it takes
on different values according to its direction; also the gluon
velocity is anisotropic. In the case of the cubic crystal the
gluon dispersion law for small momenta is however isotropic.   It
is interesting to remark that at least for gluon long wavelengths,
the essential anisotropy of the cubic structure is washed out,
while it remains  visible for the plane wave crystal structure.

\begin{center}
{\bf{Acknowledgements}} \end{center}Two of us, R.C. and G. N.,
wish to thank the DESY theory group, where this work was
completed, for the very kind hospitality.

\section*{Appendix 1: The cube symmetry group $O_h$}
The cube symmetry group $O_h$ is the finite group containing the
transformations that leave the cube (see Fig. 1) invariant. It is
defined as follows: \be O_h\ =\ O\,\times\,C_i \ee where $O$ is
the octahedron  group (see below) and $C_i=\{E,\,I\}$ is the
center of symmetry of the group, containing the neutral element
$E$ and the inversion $I$: \be E=
  \left(\matrix{1& 0 & 0\cr 0 & 1 & 0\cr 0 & 0& 1}\right),~~~~~~~
I=
  \left(\matrix{-1& 0 & 0\cr 0 & -1 & 0\cr 0 & 0& -1}\right),~
\ee i.e.\bea E&:&~~~~~
x_1\to x_1,~~~x_2\to x_2,~~~x_3\to x_3 \label{ebasic}\\
I&:&~~~~~ x_1\to -x_1,~~~x_2\to -x_2,~~~x_3\to -x_3
\label{ibasic}\ .\eea The group $O_h$ contains 48 elements,
comprising the 24 elements of the octahedron group $O$ and other
24 given by $g\cdot I$ ($g\in O$). The octahedron group is defined
by the following system of axes of symmetry:
\begin{itemize}
\item Three 4-fold axes\footnote
{For any crystal a n-fold axis is such that the crystal assumes
the same position after a rotation of an angle ${2\pi}/{n}$ around
it. Such a rotation is denoted as $C_n$; for extension  we denote
by $C_n$ not only the rotations but also the corresponding
symmetry axes.}, denoted as $C_4$ in Fig. 1;
\item four 3-fold axes, denoted as $C_3$ in Fig. 1;
\item six 2-fold axes, denoted as $C_2$ in Fig. 1.
\end{itemize}
Its 24 elements are divided into five classes:
\begin{itemize}
\item Class I, containing only $E$;
\item class II, containing three  rotations $C_4$
and three  rotations $C_4^3$  around the  three
 4-fold axes\footnote{In general
$C_n^m$ denotes a rotation of an angle
 $m\times{2\pi}/{n}$ around  a n-fold axis; clearly
  $C_n^n=1$.};
\item class III containing three rotations $C_4^2$
 three rotations of $2\times{2\pi}/{4}=\pi$ around  the three
 4-fold axes;
 \item class IV, containing four  rotations $C_3$
 and four  rotations $C_3^2$ around the four 3-fold axes;
\item class V, containing six rotations $C_2$ around the six 2-fold axes;
\end{itemize}
It is easy to construct all the 24 group elements.
 Class II comprises the  three $C_4$ rotations, i.e. rotations
 of $\pi/2$ around the three 4-fold axis:\\
  \be {\mathbf{C_4}}:~~~R_1=
  \left(\matrix{1& 0 & 0\cr 0 & 0 & 1\cr 0 & -1 & 0}\right),~~R_2=
  \left(\matrix{0& 0 & -1\cr 0 & 1 & 0\cr 1 & 0 & 0}\right),~~R_3=
  \left(\matrix{0& 1 & 0\cr -1 & 0 & 0\cr 0 & 0 & 1}\right)
    \ee
and  the three $C_4^3$ rotations: \be
{\mathbf{C_4^3}}:~~~R_1^3,~~R_2^3,~~R_3^3
      \ee

 Class III comprises the  three $C_4^2$ rotations:\\  \be
{\mathbf{C_4^2}}:~~~R_1^2,~~
  R_2^2,~~
  R_3^2
    \ee

 Class IV comprises four $C_3$ rotations:
 \be {\mathbf{C_3}}:~~~
 A_1=R_1R_2,~~~A_2=R_1A_1R_1^{-1},~~~
 A_3=R_2A_1R_2^{-1},
 ~~~A_4=A_3A_1A_3^{-1}  \ee
and four  $C_3^2$ rotations:
  \be {\mathbf{C_3^2}}:~~~
 A_1^2~,~~~A_2^2~,~~~A_3^2~,~~~A_4^2;
    \ee
finally class V comprises six $C_2$ rotations:
 \bea
 {\mathbf{C_2}}:~~~B_1&=&A_1R_2~,~~B_2=A_1B_1A_1^{-1}~,~~B_3=A_2B_1A_2^{-1}~,\cr
 B_4&=&A_3B_1A_3^{-1}~,~~~B_5=A_4B_1A_4^{-1}~,~~B_6=R_1B_1R_1^{-1}~.\eea

We finally observe that from the general theorem for finite groups
asserting that the sum of the squares of the dimension of the
representations, $n_i$, must be equal to the order of the group
(in this case 24):\be \sum_i n_i^2 =24\ee  the only possible
solution for $O$ is \be n_1=n_2=1,~~~n_3=2,~~~n_4=n_5=3.\ee

\section*{Appendix 2}

In this Appendix we evaluate the coefficients appearing in eq.
(\ref{57}): \be \beta_{\ell m}^{ij}= \sum_{k=1}^8
 \epsilon^{(k)}_i\epsilon^{(k)}_j\int\frac{d\vec v }{8\pi}v_\ell\,
 v_m\,
\delta\left\{\vec v\cdot\vec n_k-\frac{\delta \mu}{q}\right\}\ .
\ee To this end we make use of the result
 \bea
\sum_{\vec v}\,v_iv_j\,\delta\left\{\cos\theta-\frac{\delta
\mu}{q}\right\}  \, \equiv  \, \int \frac{d \vec v}{8\pi} \,
v_iv_j \,\delta\left\{\cos\theta-\frac{\delta \mu}{q}\right\} \, =
\cr = \left(\frac{\sin^2\theta_q}{8}\,
\left(\delta_{i1}\delta_{j1}+\delta_{i2}\delta_{j2}\right)
\,+\,\frac{\cos^2\theta_q}{4}\delta_{i3}\delta_{j3} \right)\ \eea

with \be \cos^2\theta_q=\left(\frac{\delta \mu}{q}\right)^2.\ee

We therefore obtain
\be \beta_{\ell m}^{ij}= \sum_{k=1}^8
R^{(k)}_{p\ell}R^{(k)}_{qm}C^{(k)}_{ij}\left(\frac{\sin^2\theta_q}{8}\delta_{pq}
\,+\,\frac{3\cos^2\theta_q-1}{8}\delta_{p3}\delta_{q3} \right) ~,
\label{bijlm}\ee where the eight matrices $R^{(k)}$ are rotation
matrices that transform the eight vectors $\vec n_k$ into the unit
vector $(0,0,1)$. They are given by \be R^{(1)}=
\left(\begin{array}{cccc}
   \frac 1{\sqrt 2}& -\,\frac 1{\sqrt 2}& 0  \\
  \frac 1{\sqrt 6} & \frac 1{\sqrt 6}&-\,{\sqrt \frac 2 3}  \\
  \frac 1{\sqrt 3} & \frac 1{\sqrt 3}&\frac 1{\sqrt 3}
\end{array}\right)\ee and, in the notations of the Appendix 1
 \bea
R^{(2)}&=&R^{(1)}R_3^{-1},~~~ R^{(3)}=R^{(1)}R_3^2,~~~
R^{(4)}=R^{(1)}R_3,\cr R^{(5)}&=&R^{(1)}B_3,~~
R^{(6)}=R^{(2)}B_4,~~ R^{(7)}=R^{(3)}B_3,~~ R^{(8)}=R^{(4)}B_4\ .
 \eea
On the other hand the eight matrices $C^{(k)}_{ij}$ are \be
C^{(k)}_{ij}= \left(\begin{array}{cccc}
   1& \eta_3^{(k)}& \eta_2^{(k)}  \\
  \eta_3^{(k)} & 1&\eta_1^{(k)}  \\
  \eta_2^{(k)} & \eta_1^{(k)}&1
\end{array}\right)_{ij}\ee
with \bea
  \eta_1^{(k)}&=&
(+1,-1,-1,+1,-1,+1,+1,-1) \cr \eta_2^{(k)}&=&
(+1,+1,-1,-1,-1,-1,+1,+1)\cr
\eta_3^{(k)}&=&(+1,-1,+1,-1,+1,-1,+1,-1)\ . \eea

Let us evaluate $\beta^{ij}_{lm}$. We notice that \be
C_{ij}^{(k)}=\delta_{ij}+\sum_{r=1,2,3}\eta_r^{(k)}\rho^r_{ij}\label{61}\ee
where \be
\rho^r_{ij}=\delta_{is}\delta_{jt}+\delta_{it}\delta_{js},~~~~r,s,t~{\rm
in~ cyclic~ order}\ee and \be \sum_{k=1,\cdots,8} \eta_r^{(k)}=0,
~~~~~\sum_{k=1,\cdots,8}\eta_r^{(k)}\eta_s^{(k)}=8\delta_{rs}\
.\ee
Using the orthogonality of the $R^{(k)}$ matrices we can evaluate
the first term of the r.h.s. of (\ref{bijlm}) immediately \be
\beta_{\ell m}^{ij} \, = \, \sin^2\theta_q\delta_{lm}\delta_{ij}+
\frac{3\cos^2\theta_q-1}{8}\sum_{k=1}^8
R^{(k)}_{3\ell}R^{(k)}_{3m}C^{(k)}_{ij} ~. \ee From the very
definition of the matrices $R^{(k)}$ we get \be
R_{ij}^{(k)}n^{(k)}_j=\delta_{i3} \to
R_{3j}^{(k)}=n^{(k)}_j=\frac{\epsilon_j^{(k)}}{\sqrt{3}}\ .\ee
Therefore \be \beta_{\ell m}^{ij} \, = \,
\sin^2\theta_q\delta_{lm}\delta_{ij}+
\frac{3\cos^2\theta_q-1}{24}\sum_{k=1}^8C^{(k)}_{lm} C^{(k)}_{ij}
~. \ee Using the expression (\ref{61}) for $C^{(k)}_{ij}$ we find
\be \beta_{\ell m}^{ij}=\frac 2 3\delta_{lm}\delta_{ij}+
\frac{3\cos^2\theta_q-1}{3}\sum_{r=1}^3\rho^r_{lm} \rho^r_{ij} ~.
\ee
\section*{Appendix 3}

For the case of a single plane wave we get the following
expression for the polarization tensor: \bea -\,\Pi^{\mu \nu}_{a
b}(p) &=&i \delta_{a b} \frac{g^2 \mu^2}{2 \pi^3} \sum_{\vec v}
  \Big[  V^{\mu}V^{\nu} \tilde I_1(p)+ \tilde V^{\mu}\tilde V^{\nu}  I_1(p)
 +\left(V^{\mu} \tilde V^{\nu} + V^{\nu} \tilde
V^{\mu}\right)I_2(p)\Big]  \cr &-&i \delta_{a b} \frac{g^2}{\pi^3}
\sum_{\vec v}
 J\,\Big(2 g^{\mu\nu}
 - \tilde V^{\mu} V^{\nu} - \tilde V^{\nu}
V^{\mu}\Big) \label{piprimo}\,,\eea where

\be I_1(p)=\int \! d^2 \ell\,  \frac{   V\cdot
 \ell\,
  V \cdot(\ell+p)}{D(\ell+p) D (\ell)}\,,\ee
\be \tilde I_1(p)=\int \! d^2 \ell\,  \frac{ \tilde  V\cdot
 \ell\,
 \tilde V \cdot(\ell+p)}{D(\ell+p) D (\ell)}\,,\ee
\be I_2(p)=\int \! d^2 \ell\,  \frac{   \Delta^2_{eff}}{D(\ell+p)
D (\ell)}\,,\ee \be J=\int_{-\mu}^{+\mu}
  d\ell_\parallel (\mu+\ell_\parallel)^2
  \int_{-\infty}^{+\infty} d\ell_0\,  \frac{\tilde  V \cdot \ell}{(2
  \mu+\tilde V \cdot \ell+i\epsilon)D(\ell)
}\,,\ee and $D (\ell) = V \cdot \ell \, \tilde V \cdot \ell -
\Delta_{eff}^2+i\epsilon$. We note that the contribution
proportional to $J$ in (\ref{piprimo}) arises from the diagram of
fig 2 b and corresponds to the lagrangian ${\cal L}_2$ in
(\ref{dirac}).

The integrals $I_1(p)$ and $\tilde I_1(p)$ are infrared divergent
in the blocking region where $\Delta_{eff}=0$. One has to
regularize them going to finite temperature and imaginary
frequencies ($\ell_0\to 2\pi i T(n+1/2)$)  (for a similar case
see \cite{rivistabeppe},  eqns. (2.193) and (2.194)). The results
that we get for the blocking and the pairing region are
respectively: \be I_1(0)=I_1(p)=\tilde I_1(0)=\tilde I_1(p)=-2\pi
i,~~I_2(0)=I_2(p)=0,~~J=\frac{\pi i}2\mu^2\ee and \be
I_1(0)=I_1(p)=\tilde I_1(0)=\tilde I_1(p)=-\pi i,~~ I_2(0)=-\pi
i,~~J=\frac{\pi i}2\mu^2\,.\ee From these we can easily compute
the Debye and Meissner masses using \bea -\,\Pi^{\mu \nu}_{a
b}(0)\Big |_{blocking} &=& \delta_{a b} \frac{g^2 \mu^2}{2 \pi^2}
\sum_{\vec v;\, blocking}
  \Big[  2\left(V^{\mu}V^{\nu}  + \tilde V^{\mu}\tilde V^{\nu}
  \right)
 \cr&+&2 g^{\mu\nu}
 - \tilde V^{\mu} V^{\nu} - \tilde V^{\nu}
V^{\mu}\Big] \eea
 \bea -\,\Pi^{\mu \nu}_{a b}(0)\Big |_{pairing}
&=& \delta_{a b} \frac{g^2 \mu^2}{2 \pi^2} \sum_{\vec v;\,
pairing}
  \Big[  V^{\mu}V^{\nu}  + \tilde V^{\mu}\tilde V^{\nu}
   \cr&+&2 g^{\mu\nu}
 - 2\left(\tilde V^{\mu} V^{\nu} + \tilde V^{\nu}
V^{\mu}\right)\Big]\,. \eea We get therefore the vanishing of the
Meissner mass \be\Pi_{ab}^{ij}(0)=0\,.\ee On the other hand we
get a non-vanishing Debye screening mass from the blocking region:
\be -\,\Pi^{00}_{a b}(0)\ =\ 2 \delta_{ab} \frac{g^2 \mu^2}{\pi^2}
\sum_{\vec v;\, blocking} = \delta_{ab} \frac{g^2 \mu^2}{2 \pi^2}
\int_{blocking} d \cos \theta\, ~, \label{blockmass} \ee where
\be \int_{blocking} d \cos \theta=\int_{-1}^{\cos\theta_a} d \cos
\theta+\int_{\cos\theta_b}^1 d \cos \theta\,.\ee The domain of
integration of the blocking region has been obtained by the
condition \be R|h(\cos\theta)| > \frac\pi 2\ ,
\label{blockdomain} \ee which gives, using the approximate
expression for $h(\cos\theta)$ (\ref{hdix}), and for $R>\pi/2$:
\be \cos\theta_a=\frac{\cos\theta_q}{\dd 1+\frac{\pi}{2R}},~~~~~
\cos\theta_b=min\left\{1,\,\frac{\cos\theta_q}{\dd
1-\frac{\pi}{2R}}\right\}\, , \ee where
$\cos\theta_q=\delta\mu/q$.
 From the equations (\ref{blockmass}) and
(\ref{blockdomain}) one gets the Debye mass:  \be m_D = \frac{g
\mu}{\pi} \sqrt{ 1 + \frac{\cos \theta_a-\cos\theta_b}{2} }\,.\ee
These results can be easily generalized to the case of the cube.

\end{document}